\documentclass[iop]{emulateapj} 

\usepackage{color,graphicx}
\usepackage{amsmath,amssymb}
\usepackage{tabularx}

\newcommand{\iso}[2]{\hbox{${}^{#1}{\rm #2}$}}
\newcommand{\Msun}{\ensuremath{{M}_{\sun}}}
\newcommand{\Lsun}{\ensuremath{{\rm L_{\sun}}}}

\shorttitle{Stellar yields from metal-rich models}
\shortauthors{Karakas \& Lugaro}

\begin{document}


\title{Stellar yields from metal-rich asymptotic giant branch models}


\author{Amanda I. Karakas\altaffilmark{1,3} and Maria Lugaro\altaffilmark{2,3}}
\email{amanda.karakas@anu.edu.au}


\altaffiltext{1}{Research School of Astronomy and Astrophysics, 
  Australian National University, Canberra, ACT 2611, Australia}
\altaffiltext{2}{Konkoly Observatory, Research Centre for Astronomy and Earth Sciences,
Hungarian Academy of Sciences, H-1121 Budapest, Hungary}
\altaffiltext{3}{Monash Centre for Astrophysics, School of Physics and Astronomy,
Monash University, VIC 3800, Australia}


\begin{abstract}
We present new theoretical stellar yields and surface abundances for three grids
of metal-rich asymptotic giant branch (AGB) models. Post-processing 
nucleosynthesis results are presented for stellar models with initial masses between 
1$\Msun$ and 7.5$\Msun$ for $Z=0.007$, and 1$\Msun$ and 8$\Msun$ for $Z=0.014$ (solar) 
and $Z=0.03$. We include stellar surface abundances as a function of thermal pulse
on the AGB for elements from C to Bi and for a selection of isotopic ratios for  
elements up to Fe and Ni (e.g., \iso{12}C/\iso{13}C), which can be  
obtained from observations of molecules in stars and from the laboratory analysis of 
meteoritic stardust grains. Ratios of elemental abundances of He/H, C/O, and N/O are 
also included, which are useful for direct comparison to observations of AGB stars and 
their progeny including planetary nebulae.
The integrated elemental stellar yields are presented for each model in the
grid for hydrogen, helium and all stable elements from C to Bi. Yields of Li
are also included for intermediate-mass models with hot bottom burning.
We present the first $slow$ neutron-capture ($s$-process) yields for super-solar 
metallicity AGB stars with $Z=0.03$, and the first complete $s$-process yields 
for models more massive than 6$\Msun$ at all three metallicities.
\end{abstract}






\section{Introduction}

Theoretical stellar nucleosynthesis calculations are an important dataset 
for the interpretation of chemical abundances that are derived from spectra of
stars and gaseous regions in galaxies. When the abundances are from samples
of old, low-mass stars they allow us to disentangle the processes of galaxy 
formation and evolution; the study of Galactic archeology \citep{freeman02}.
In this framework, stellar abundances are compared to theoretical predictions from
chemical evolution models, which require as input the chemical
yields from stars under the assumption that the yields from previous generations of
stars have contributed to the build up of elements over time 
\citep{romano10,kobayashi11a,kobayashi11b,shingles14,molla15}

Theoretical nucleosynthesis calculations are also essential for a direct comparison 
between predicted stellar abundances and observations. This is especially so for evolved 
stars that are on the red giant branch (RGB) and the asymptotic giant branch (AGB), or have 
evolved to become post-AGB stars and planetary nebulae (PNe). In this case, comparison 
between theory and observation can provide insights into our understanding of stellar 
astrophysics. If we fail to explain the amount of neutron-capture elements on the surface 
of a post-AGB star, for example, clearly the model needs improving \citep[e.g.,][]{desmedt12}. 
Furthermore, detailed stellar nucleosynthesis calculations providing isotopic 
abundances are needed for comparison to direct observations of isotopic ratios 
from molecular lines \citep[e.g.,][]{kahane00,lederer09,milam09,fonfria15}. 
They also represent the key to interpret the composition of stable 
and radioactive isotopes in meteoritic components, such as stardust grains 
\citep{zinner14} and calcium-aluminum inclusion \citep[e.g.,][]{akram13}, as well as 
meteoritic leachates and whole rocks \citep[e.g.,][]{dauphas02,burkhardt15,akram15}, 
with implications on the origin of cosmic dust and on the formation of the Solar System.

Low and intermediate-mass stars cover a range in mass from 0.8 -- 8$\Msun$, depending on
metallicity \citep[see Fig. 1 from][]{karakas14dawes}. Stars with initial 
masses in this range will evolve through core hydrogen and helium burning before 
ascending the AGB \citep{busso99,herwig05,karakas14dawes}.
It is during the AGB phase that the richest nucleosynthesis occurs, driven
by He-shell instabilities. These instabilities or thermal pulses (TP) may result in
mixing between the H-exhausted core and the envelope; this is known
as third dredge up (TDU). The TDU will alter the composition of the envelope by
bringing the products of He-shell burning and the elements produced by the
$slow$ neutron capture process (the $s$-process) to the stellar surface. 

Low-mass AGB stars with initial masses $M \lesssim 4\Msun$ have surface compositions 
and stellar yields characterized by enrichments in carbon, nitrogen, fluorine, and 
$s$-process elements \citep[e.g.,][]{busso01,karakas07a,cristallo09,weiss09}.
In contrast, intermediate-mass AGB stars with initial masses $M\gtrsim 4\Msun$ experience 
both the second dredge-up (SDU) during the early AGB, which results in large increases in 
helium and nitrogen, and hot bottom burning (HBB), the process by which 
the base of the envelope becomes hot enough for proton-capture
nucleosynthesis \citep{karakas03b,ventura13}. The surface chemistry of intermediate-mass 
stars is thus characterized by proton capture nucleosynthesis, perhaps with some  
contribution from He-shell burning and the $s$-process \citep{karakas12}. 

\citet{karakas14dawes} reviewed the available stellar yields from AGB models. 
The most significant gaps were found for low-metallicity AGB models and for the yields of
$s$-process elements. For metallicities around solar, the only set of 
tabulated stellar yields of $s$-process elements are those by \citet{cristallo15}, which
are available on the FRUITY on-line database\footnote{FUll-Network Repository of Updated 
Isotopic Tables \& Yields: http://fruity.oa-teramo.inaf.it/}.
The NuGrid collaboration is also in the process of publishing yields for AGB stars 
of metallicity around solar ($Z=0.02$ and 0.01), for a limited range of masses
\citep[e.g.,][]{pignatari13}. No models exist for metallicities higher than $Z = 0.02$. 

Stellar yields are known to be highly uncertain and dependent on the
model assumptions used in the stellar evolutionary (and post-processing if used) 
calculations \citep{ventura05a,ventura05b,stancliffe07a,karakas10a}. 
These uncertainties manifest into uncertainties in chemical evolution
studies \citep{romano10}. This means that yields by different groups vary as a consequence
of assumptions about the treatment of convection and convective borders and the
adopted mass-loss rates. For that reason, it is essential to provide yields
from different stellar evolution codes in order to understand what elements
are most affected by stellar modelling uncertainties. 

Comparing yield sets is useful because it can reveal how different
choices in the input physics affects the yields. However, it does not provide
an indication of the reliability of any set of stellar yields. Until we are 
able to constrain how many thermal pulses an AGB star of a given initial mass and 
metallicity is expected to go through, the stellar yields will remain uncertain. 
Independent observational tests are the most reliable method to test the validity 
of any set of stellar models. Such a comparison will be the focus of follow-up  
studies, where we confront our predictions with available observational data
for AGB stars, PNe and post-AGB stars, and presolar grains.

In \citet{karakas14b} we provided stellar evolutionary tracks for an updated
set of low and intermediate-mass stellar models between $1-8\Msun$\footnote{we did not
provide evolutionary tracks for the post-AGB and white dwarf phases, we refer
to the recent evolutionary calculations by, e.g., \citet{miller15}.}.
In particular, we included models of solar metallicity ($Z=0.014$), super solar ($Z=0.03$)
and a factor of two below solar ($Z=0.007$). \citet{karakas14b} examined the
effect of helium enrichment on the production of carbon stars. It was found that 
modest ($\Delta Y \approx 0.05 - 0.1$) increases in helium abundance above the
canonical value inhibits carbon star production. This because
less He-intershell material is dredged to the surface, which also means that the
stellar yields of other elements (e.g., $s$-process elements in particular) will
be reduced.  This has been shown to be the case in low-metallicity AGB models 
\citep{karakas14,shingles15}. 

In this study we aim to provide theoretical nucleosynthesis predictions including
$s$-process elements from the grid of stellar evolutionary models from 
\citet{karakas14b} with a canonical helium composition.  For the first time,
we include surface abundances and stellar yields for masses 
up to the CO core limit ($\approx 8\Msun$, see \S\ref{sec:models}) and models 
of super-solar AGB stars of $Z=0.03$. We also aim to provide abundances in a form 
that is useful to the AGB, post-AGB, and PNe communities, to allow for a direct 
comparison to abundances of these objects.

Section~\ref{sec:models} discusses the stellar evolutionary models used as
input into our post-processing nucleosynthesis calculations, while our 
nucleosynthesis results are summarized in Section~\ref{sec:abund}. 
We finish with a discussion and conclusion.

\section{The Stellar Models} \label{sec:models}

For the post-processing nucleosynthesis calculations we use as input the 
stellar evolutionary models described in \citet{karakas14b}. 
We consider here only the models with a canonical helium composition, which is
$Y=0.26$ for $Z=0.007$, $Y = 0.28$ for $Z=0.014$, and $Y=0.30$ for $Z=0.03$. 
The grids includes models of solar metallicity, defined here to be $Z=0.014$ 
\citep[based on the solar abundances from][]{asplund09},
and a factor of two above and below solar: $Z =0.007$ and $Z=0.03$. 

While we refer to \citet{karakas14b} for the full details of the input physics 
and the numerical method, we remind the reader of the input physics most relevant
to the stellar nucleosynthesis.  No mass-loss is used on the RGB and we refer to 
\citet{karakas14b} for a justification of this choice. We use the \citet{vw93} mass-loss
rate on the AGB phase. We use the Mixing-length Theory of convection, with a mixing-length
parameter $\alpha = 1.86$ and assume instantaneous mixing in convective regions. 
No convective overshoot is included in the calculations prior to the AGB. Dealing
with the borders between radiative and convective regions in stellar interiors is a major
uncertainty.  We implement an algorithm to try to search for a neutrally stable 
point from the formal Schwarzschild boundary as described by \citet{lattanzio86}.
This method has been shown to increase the efficiency of third dredge-up, at least in 
intermediate-mass models of $\approx 5\Msun$ \citep{frost96}, but not in lower mass 
models close to the minimum mass for carbon stars \citep{kamath12}.

The models cover the complete range of AGB masses from 1$\Msun$ to the upper limit for producing
 a CO core, which is 8$\Msun$ for $Z=0.014$ and $Z=0.03$ and 7$\Msun$ for $Z = 0.007$. 
The 8$\Msun$ models (and the 7$\Msun$ at $Z=0.007$) produce a hybrid CO(Ne) core, and
experience off-centre carbon flashes but the temperature in the core is not high enough to 
ignite a carbon flame that reaches the centre \citep[e.g., as described by][]{siess06}. 
Above these masses, stars will become ONe super-AGB stars or neutron stars \citep{doherty15}.

We supplement the calculations in \citet{karakas14b} with  extra stellar
evolutionary model calculations such that we have a grid of models with a mass spacing
of $\Delta M = 0.25\Msun$ up to 5$\Msun$; above that mass we assume the same grid as
described in \citet{karakas14b}. The new models have masses: $M=2.75, 3.25, 3.75,
4.25, 4.75\Msun$ for the metallicities where we did not provide these masses before.
We also include a 7.5$\Msun$, $Z = 0.007$ model, which evolves through complete core C burning
before ascending the AGB as a ONe-core super-AGB star \citep[e.g.,][]{doherty14a}.
This is to make sure that we have a fine enough mass grid such that we do not miss 
any important non-linear behaviour in the stellar yields. The same stellar evolutionary 
code and input physics were adopted for the new calculations for consistency.

The theoretical minimum initial mass for producing a solar metallicity carbon-rich star in 
\citet{karakas14b} is 2$\Msun$. This is likely above the observational limit of 
$\approx 1.5\Msun$, which is derived directly from observations of C stars in binary 
systems and open clusters \citep{groen95}, although uncertainties are large and statistics are low.
A mass of $\approx 1.5\Msun$ is also derived by comparison of theoretical models to observationally
derived carbon-star luminosity functions in the Galaxy and Magellanic Clouds 
\citep[e.g.,][]{groen95,marigo99,stancliffe05a,cristallo11}. \citet{groen95} estimate
the minimum mass for solar metallicity C-stars to lie between 1.5--1.6$\Msun$,
where their $Z_{\rm solar} = 0.02$.  Using updated observational data, \citet{cristallo11} provide 
a minimum mass as a function of metallicity: for $Z=0.02$ the minimum mass is 1.5$\Msun$ and 
for $Z=0.01$ the minimum mass is 1.4$\Msun$. 

It should be stressed that the minimum mass derived from these studies is dependent upon 
the underlying assumptions in the theoretical calculations as well as the uncertainties 
in the photometry. This can be highlighted by examining the results from the study by 
\citet{kalirai14}, who use white dwarfs in open clusters to study the core mass growth 
and initial-final mass relation. These authors come to conclusion that no third dredge-up 
takes place in stars less than 2$\Msun$ for a metallicity $Z=0.02$ (which they assume 
is slightly super-solar). The \citet{kalirai14} results are also model dependent but 
depend on a different code \citep{marigo13}. GAIA will provide much help here by providing 
the distances and hence luminosities to well known, bright C-rich stars in the Galaxy. 
Until then, the minimum mass for C-stars in the Galaxy is not accurately known and probably 
lies somewhere between 1.4--2$\Msun$.

Motivated by these uncertainties, we include convective overshoot at the base of 
the convective envelope such that a selection of low-mass AGB models also become C-rich.
These are  the $M=1.5, 1.75\Msun$, $Z=0.007, 0.014$ models, and the $M=2.5, 2.75, 3\Msun$, 
$Z=0.03$ models.  For the $Z=0.03$ models we do not have observational clues as to the
minimum mass for C-stars but it is likely $\gtrsim 2\Msun$ \citep{kalirai14}. 
Given the uncertainty in the minimum mass for C-stars in the Galaxy, we provide surface 
abundances and yields from both calculations: those with overshoot and those without. 
The prescription  we use to include overshoot is the same as used by \citet{karakas10a} 
and \citet{kamath12}, and extends the base of the envelope by $N_{\rm ov}$ pressure scale 
heights during dredge-up.  In order for the masses considered here to become C-rich, we use 
$N_{\rm ov} \le 3$, with the specific values used given in the footnotes of 
Table~\ref{tab:models}. These values are consistent with what was found by 
\citet{kamath12} in order to reproduce the observed M/C transition luminosity of 
AGB stars in three Magellanic Cloud  clusters. This convective overshoot prescription
is not applied to models above 1.75$\Msun$ for $Z \le 0.014$ and above 3$\Msun$ for $Z=0.03$
so does not affect intermediate-mass models with HBB.

\begin{table*}
\begin{center}
\caption{Stellar models calculated in addition to those in \citet{karakas14b}. 
The luminosity is in the format $n(m)$ where $= n \times 10^{m}\Lsun$.}
\label{tab:models}
\begin{tabular}{lcccccccccccc} \hline \hline
Mass & SDU & HBB & TDU & \#TP & $\lambda_{\rm max}$ & $M_{\rm c}(1)$
 & $T_{\rm bce}^{\rm max}$ & $L_{\rm agb}^{\rm max}$ & 
$\tau_{\rm stellar}$ & $\tau_{\rm agb}$ & $\tau_{\rm tpagb}$ \\
($\Msun$) &  &     &     &    &     & ($\Msun$) & (MK) & ($\Lsun$) & 
 (Myr) & (Myr) & (Myr)  \\ \hline
\multicolumn{12}{c}{$Z=0.007$, $Y=0.26$ models.} \\ \hline
1.50 & No & No & Yes$^{\rm a}$ & 18 & 0.31 & 0.548 & 3.77 & 8.39(3) &
2451 & 18.67 & 2.057 \\
1.75 & No & No & Yes$^{\rm a}$ & 17 & 0.51 & 0.556 & 4.13 & 8.68(3) &
1535 & 17.05 & 1.836 \\
2.75 & No & No & Yes & 25 & 0.80 & 0.589 & 5.11 & 1.22(4) &
509.6 & 13.09 & 2.086 \\
3.25 & No & No & Yes & 22 & 0.92 & 0.698 & 11.3 & 1.54(4) &
313.4 & 6.115 & 0.920 \\
3.75 & No & No & Yes & 22 & 0.97 & 0.786 & 26.8 & 1.94(4) &
212.2 & 3.964 & 0.503 \\
4.75 & Yes & Yes & Yes & 55 & 0.95 & 0.865 & 80.5 & 3.43(4) &
118.0 & 2.076 & 0.528 \\  
7.50 & Yes & Yes & Yes & 87 & 0.83 & 1.107 & 109 & 7.32(4) &
42.6 & 0.383 & 0.075 \\ \hline
\multicolumn{12}{c}{$Z=0.014$, $Y=0.28$ models.} \\ \hline
1.50 & No & No & Yes$^{\rm b}$ & 16 & 0.51 & 0.552 & 6.24 & 7.24(3) &
2882 & 18.26 & 1.512 \\
1.75 & No & No & Yes$^{\rm c}$ & 20 & 0.50 & 0.554 & 5.43 & 8.73(3) &
1755 & 18.61 & 1.736 \\
3.75 & No & No & Yes   & 23 & 0.95 & 0.740 & 18.7 & 3.10(4) &  
229.4 & 5.129 & 0.620 \\ 
4.25 & Yes & Yes & Yes & 31 & 0.96 & 0.840 & 61.9 & 2.67(4) &
162.9 & 3.176 & 0.377 \\ 
4.75 & Yes & Yes & Yes & 35 & 0.95 & 0.855 & 68.8 & 2.90(4) & 
121.9 & 2.322 & 0.339 \\ \hline 
\multicolumn{12}{c}{$Z=0.03$, $Y=0.30$ models.} \\ \hline
2.50 & No & No & Yes$^{\rm d}$ & 30 & 0.81 & 0.547 & 8.82 & 1.08(4) &
914.8 & 26.44 & 2.211 \\
2.75 & No & No & Yes$^{\rm c}$ & 33 & 0.80 & 0.559 & 9.03 & 1.21(4) &
695.3 & 21.17 & 2.076 \\
3.00 & No & No & Yes$^{\rm a}$ & 33 & 0.81 & 0.580 & 8.24 & 1.32(4) &
532.4 & 16.90 & 1.719 \\
\hline \hline
\end{tabular}
\\
(a) $N_{\rm ov} = 1$, see text for details.\\
(b) $N_{\rm ov} = 3$.\\
(c) $N_{\rm ov} = 2$.\\
(d) $N_{\rm ov} = 2.5$.
\medskip\\
\end{center}
\end{table*}

Table~\ref{tab:models} shows some the relevant properties of the additional AGB
models calculated, including the initial stellar mass, whether the models experience 
second dredge up (SDU), hot bottom burning (HBB), and third dredge-up (TDU). For the cases
where we include overshoot we provide a footnote with the $N_{\rm ov}$ parameter used.
We include the total number of thermal pulses (\#TP), the maximum value of the dredge-up 
efficiency parameter\footnote{$\lambda = \Delta M_{\rm dredge}/\Delta M_{\rm c}$,
where $\Delta M_{\rm dredge}$ is the amount of material dredged up, and $\Delta M_{\rm c}$
is the core mass growth during the preceding interpulse phase, see \citet{karakas02}.},
$\lambda_{\rm max}$, the core mass at the first thermal pulse, $M_{\rm c}(1)$, 
the maximum temperature at the base of the convective envelope,  $T_{\rm bce}^{\rm max}$, 
and the maximum surface luminosity during the AGB, $L_{\rm agb}^{\rm max}$.
For low-mass AGB stars $\lesssim 4\Msun$ 
the maximum luminosity occurs at the tip of the AGB 
and is generally higher than the luminosity when the star becomes C-rich \citep{kamath12}.
For models with HBB, the maximum luminosity occurs before the tip of the AGB, when HBB
is active.  We also provide the total stellar lifetime, $\tau_{\rm stellar}$, 
the AGB lifetime, $\tau_{\rm agb}$, and the thermally-pulsing 
AGB lifetime, $\tau_{\rm tpagb}$. 
Units are Myr for lifetimes, solar units for masses and luminosities,  and 
MK ($10^{6}$ K) for temperatures. 

\begin{figure}
\begin{center}
\includegraphics[width=0.95\columnwidth]{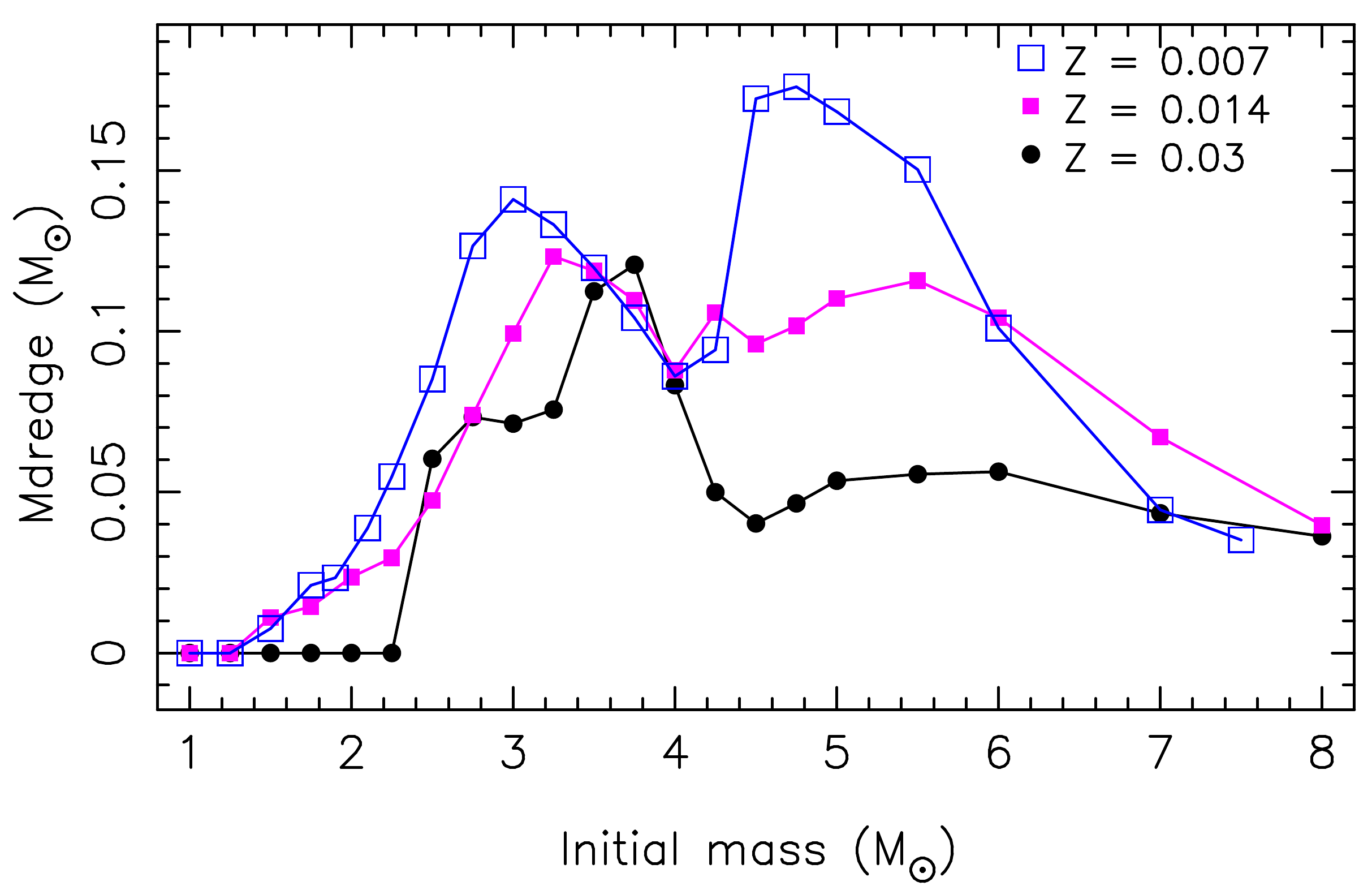}
 \caption{The total amount of mass dredged-up to the envelope by TDU for models
of $Z = 0.007$, $Z=0.014$, and $Z=0.03$.
\label{mdredge}}
\end{center}
\end{figure}

In Fig.~\ref{mdredge} we show the total amount of mass dredged into the envelope
by TDU, $M_{\rm dredge}$. In this figure and from now on we present and 
discuss models with convective overshoot
that become C-rich (e.g., the models from Table~\ref{tab:models}) in preference 
to models of the same mass without overshoot, unless specified. 
Fig.~\ref{mdredge} shows that the amount of material dredged up increases with decreasing 
metallicity, for a given mass. The behaviour of $M_{\rm dredge}$ with initial stellar mass is 
interesting: the most massive models above 7$\Msun$ show smaller values of 
$M_{\rm dredge}$ as a consequence of their thinner He-intershell regions. However, 
around 4--6$\Msun$ there is an increase in $M_{\rm dredge}$ as a consequence of 
the models experiencing many more thermal pulses relative to their lower mass 
counterparts. This means that the total yield of C$+$N for example will be higher 
in these models. Fig.~\ref{mdredge} will be useful later when discussing the behaviour 
of the yields and final surface abundances as a function of mass and metallicity.

\subsection{Post-processing nucleosynthesis calculations}

Post-processing calculations have been performed on all of the canonical helium composition
models from \citet{karakas14b} and all of the new models in Table~\ref{tab:models}.
  
The numerical method used for the post-processing numerical calculations is 
the same as described in previous studies including \citet{lugaro12}, 
\citet{fishlock14b} and \citet{shingles15}. 
The reaction rates are from the JINA reaclib database, as for May 2012, except for the 
neutron-capture cross section of the Zr isotopes, which were updated by \citet{lugaro14a}.
The one major difference here is that we have updated the nuclear network used in the 
calculations to include more nuclear species, 328 instead of the 
previous 320, and the description of the temperature dependence of $\beta$-decay rates 
for a number of unstable isotopes. This was done to follow in more detail 
the behaviour of a selection of long lived radioactive isotopes and of 
branching points on the $s$-process path: Specifically $^{107}$Pd, $^{127,129}$I, and 
$^{181,182}$Hf from \citet{lugaro14b}, as well as 
at $^{134,135,136,137}$Cs, $^{154,155}$Eu, and $^{160}$Tb. 

Here we present elemental abundances for elements heavier than Fe and Ni,
for which the treatment of branching points does not have a major effect on the 
results\footnote{except for the specific cases of Rb and  
and Cs, which are affected by the branching 
points located at unstable nuclei on the $s$-process path at $^{85}$Kr, $^{86}$Rb, and $^{134}$Cs
(all treated correctly in our network) and of Tl, which can be mildly 
affected by the branching points
at $^{203}$Hg and $^{204}$Tl (for which we have not yet implemented the predicted temperature 
dependence of the decay rates).}. 
Results for isotopic ratios of elements heavier than iron can however be strongly 
affected by branching points. These are not presented here, and will be discussed instead 
in dedicated, forthcoming papers aimed at comparing our results 
with the isotopic compositions observed in meteoritic inclusions and stardust grains.

\subsection{The inclusion of \iso{13}C pockets} \label{sec:c13pocket}

\begin{table*}
\begin{center}
\caption{Choice of $M_{\rm mix}$ for stellar models in different mass ranges. We also
show additional stellar models (mass in $\Msun$, metallicity) calculated 
with different values of $M_{\rm mix}$ for metallicities other than $Z=0.014$.}
\label{tab:pockets}
  \vspace{1mm}
\begin{tabular}{lcccc} \hline \hline
$M_{\rm mix}/\Msun$ = & 0 & $1 \times 10^{-4}$ & $1 \times 10^{-3}$ & $2 \times 10^{-3}$ \\
\hline
standard & M$\ge 5 \Msun$ & $4 < M < 5 \Msun$ & $3 < M \leq 4 \Msun$ & M$\leq 3 \Msun$ \\ 
\hline
extra models & (4.25, 0.03)   & (5, 0.03) & (3.0, 0.007) & (3.25, 0.03) \\
             & (4.5, 0.007)   & (4, 0.007) & (4.25, 0.03) &  \\
             & (4.75, 0.007)  &  & (4.5, 0.03)  &  \\
\hline \hline
\end{tabular}
\medskip\\
\end{center}
\end{table*}

To match the observations that show that AGB stars, their progeny, and their companions are 
enriched in the abundances of the $s$-process elements by up to 1 dex at  
solar metallicity \citep[see e.g.,][]{busso01,abia02}, a large number of neutrons 
are needed to be released in the intershell via ($\alpha$,n) reactions. The main source of neutrons 
is the \iso{13}C($\alpha$,n)\iso{16}O reaction, which is activated at relatively low
temperatures of $\sim$ 90 MK. However, CN cycling does not leave enough \iso{13}C nuclei in 
the He-intershell. The standard solution to this problem is to assume that some partial 
mixing occurs between the  convective H-rich envelope and the intershell at the deepest 
extent of each TDU,  so that the protons are captured by \iso{12}C to produce 
a region rich in \iso{13}C, the so-called \iso{13}C ``pocket''. 
The inclusion of \iso{13}C pockets in theoretical calculations of AGB stars is
one of the most significant uncertainties affecting predictions of the $s$ process 
\citep[see discussion in][]{busso99,herwig05,karakas14dawes}. Here, we adopt the 
same techniques we have applied before in, e.g., \citet{fishlock14b}.

Our method is to insert protons at the deepest extent of each TDU
episode in the post-processing calculations. The protons are partially mixed 
over a mass extent in the intershell denoted by $M_{\rm mix}$, using an exponentially 
declining profile such that at the base of the envelope the proton abundance is 
$X_{\rm p} \approx 0.7$ (i.e., the envelope hydrogen abundance) and in
the intershell, at $M_{\rm mix}$  below the base of the convective envelope, 
the hydrogen abundance is $X_{\rm p} = 1\times 10^{-4}$. 
Below this point in mass $X_{\rm p} = 0$. 

Our method differs from that of \citet{cristallo15}, who include time-dependent 
convective overshoot in their models at the base of the envelope which, at the deepest 
extent of each TDU episode, leads to the mixing of protons into the intershell that produces 
the \iso{13}C pocket \citep{cristallo09}. This method is more self-consistent than ours, 
since we insert directly the proton abundance profile, rather than the mixing process that 
leads to it. However, our $s$-process results for AGB stars of low mass, where \iso{13}C 
is the main neutron source, are reasonably similar to those of \citet{cristallo15}, as 
discussed by \citet{lugaro12}, \citet{fishlock14b}, and in Section~\ref{sec:compare}. 
This demonstrates that our parametric approach is a good reproduction of the 
self-consistent model of time-dependent convective overshoot. While our method 
ignores any feedback from the partially CN cycling of the protons on the structure 
it has the advantage that it allows us to easily adjust the $M_{\rm mix}$ parameter 
and the proton profile and study the effect of their variations on AGB 
nucleosynthesis \citep[e.g.,][]{lugaro14a,lugaro15}.

Evidence for variations in the quantity and distribution of $s$-process elements 
come from a number
of observational sources including post-AGB stars \citep[e.g.,][]{axel07b,desmedt12}, 
AGB stars and chemically peculiar stars that show the chemical signature of mass transfer 
from low-mass AGB stars 
\citep[e.g., Ba and CH stars, carbon enhanced metal-poor stars; ][]{busso01,bisterzo11,lugaro12} 
as well as pre-solar grains \citep{lugaro03b,lugaro14a}. 
These may be due to variation in the size of the \iso{13}C pockets and/or in the 
profile of the proton abundance leading to their formation. However, they cannot 
be theoretically derived from first principles because we do not know the physical 
mechanism responsible for producing the partial mixing of protons into the
top layers of the He-intershell. We refer to discussions in \citet{herwig05} and 
\citet{cristallo09}. Furthermore, stellar rotation has been also shown to affect the 
quantity and the distribution of the $s$-process elements produced in the \iso{13}C pocket
\citep{herwig03,siess04,piersanti13}. In this context, however, 
the effect of magnetic fields has not been investigated yet.

We include \iso{13}C pockets in all the low-mass AGB models below $4.5\Msun$ that experience TDU. 
We do not include rotation or magnetic fields in our models and we keep the same exponential 
proton profile described above in all models and for all \iso{13}C pockets, however, we experiment with
changing the value of the $M_{\rm mix}$ parameter to produce larger or smaller pockets, in terms of their 
extension in mass. 
The value of $M_{\rm mix}$ was chosen as function of the stellar mass, with our standard choices
for each mass range listed in Table~\ref{tab:pockets}. Additional models are calculated with
different values of  $M_{\rm mix}$; in Table~\ref{tab:z014pmz} we show the entire range of 
stellar nucleosynthesis models calculated for $Z=0.014$, which are the most extensive. 
In Table~\ref{tab:pockets} we also provide a list of the additional models calculated for 
the other metallicities.

\begin{table*}
 \begin{center}
  \caption{The stellar nucleosynthesis models calculated for $Z=0.014$: 
A tick (\checkmark) shows the the size of $M_{\rm mix}$ used in the calculations. 
The [ST] label indicates the cases with the standard choice for each model with TDU.}
 \label{tab:z014pmz}
  \vspace{1mm}
   \begin{tabular}{cccccc}
   \tableline\tableline
 $M_{\rm mix}/\Msun=$ & 0  & $1\times 10^{-4}$ & $1\times 10^{-3}$ &
 $2 \times 10^{-3}$ & $4 \times 10^{-3}$  \\ 
\tableline
 Stellar Mass ($\Msun$) &  &   &   &  & \\
\tableline
 1.00 & \checkmark &  &  &    &     \\
 1.25 & \checkmark &  &  &    &      \\
 1.50 & \checkmark$^{\rm a}$ &  &  & \checkmark$^{\rm b}$\,[ST]  &  \\
 1.75 & \checkmark$^{\rm a}$ &  &  & \checkmark$^{\rm b}$\,[ST] &  \\
 2.00 &            & \checkmark & \checkmark  & \checkmark\,[ST] & \checkmark \\
 2.25 &            &  &  & \checkmark\,[ST]   &  \\
 2.50 &            &  &  & \checkmark\,[ST]   &  \\
 2.75 &            &  &  & \checkmark\,[ST]   &  \\ 
 3.00 &            & \checkmark & \checkmark & \checkmark\,[ST] &   \\
 3.25 &            &  & \checkmark\,[ST] & \checkmark &   \\
 3.50 &            &  & \checkmark\,[ST] &  &  \\
 3.75 &            &  & \checkmark\,[ST] &  &  \\
 4.00 &            & \checkmark & \checkmark\,[ST] &  &  \\
 4.25 &            & \checkmark\,[ST]  & \checkmark &  &  \\
 4.50 & \checkmark & \checkmark\,[ST]  & \checkmark &  &  \\
 4.75 & \checkmark & \checkmark\,[ST]  & &  &   \\
 5.00 & \checkmark\,[ST] & \checkmark  & &  &   \\
 5.50 & \checkmark\,[ST] &  &  &    &     \\
 6.00 & \checkmark\,[ST] &  &  &    &     \\
 7.00 & \checkmark\,[ST] &  &  &    &     \\
 8.00 & \checkmark\,[ST] &  &  &    &     \\
\tableline \tableline
  \end{tabular} 
\\
(a) For the model without overshoot.\\
(b) For the model with overshoot.
\\
 \end{center}
\end{table*}

For stars of initial mass up to and equal to 3$\Msun$ we used as the standard choice $M_{\rm mix} =
2\times 10^{-3}\Msun$, which results in a \iso{13}C pocket mass typically about 1/10$^{\rm th}$ of the
mass of the He-rich intershell. 
This value of $M_{\rm mix}$ is required to match the strong observational constraint that AGB stars 
of metallicity close to solar are enhanced in $s$-process elements by up to 1 dex, as demonstrated 
previously by, e.g., \citet{gallino98}. Similar considerations were also used by \citet{cristallo09} 
to calibrate the value of the free parameter $\beta$ that controls the exponential decay of the 
velocity below the border of the convective envelope in their time-dependent description of overshoot.

For stars of initial mass between 3 and (including) 4$\Msun$ we used as
standard choice $M_{\rm mix} = 10^{-3}\Msun$. This is because the mass of the intershell decreases
and calculations including hydrodynamical overshoot indicate that $M_{\rm mix}$ also follows such
a decrease \citep{cristallo09}.   For masses between 4.25 and 5$\Msun$ we set $M_{\rm mix} = 10^{-4}\Msun$
as our standard choice, an order of magnitude smaller than for the lower masses, again
following the shrinking of the mass of the intershell. An exception is made for the 
4.75$\Msun$, $Z = 0.007$ model, which we assume does not form \iso{13}C pockets. This is 
motivated by the strong HBB experienced by this model, with temperatures reaching over 75~MK 
at the base of the envelope

For intermediate-mass AGB models above 5$\Msun$ we do not include a \iso{13}C pocket,
following indications from theory \citep{goriely04} and observations \citep{garcia13} that the
\iso{13}C pocket is not present. Instead in these models the $s$ process proceeds through
activation of the \iso{22}Ne($\alpha$,n)\iso{25}Mg reaction inside the TP, which requires
temperatures in excess of 300~MK \citep{vanraai12,karakas12}.

\section{Surface abundances during the AGB} \label{sec:abund}

Here we present a summary of the results from the post-processing nucleosynthesis 
calculations, starting with the surface abundances. 

\subsection{The surface abundance data tables} 

We provide three sets of tables, one for each metallicity: 1) the elemental surface 
abundances as a function of thermal pulse number for each ($M, Z$) combination; 2) 
the isotopic ratios of the elements up to Ni 
as a function of thermal pulse number for each ($M, Z$) combination;
and 3) the integrated elemental yields. In this section we describe the contents of the first
two tables (surface abundances and isotopic ratios), with the yield tables described in
Section~\ref{sec:yields}. 

For the ($M, Z$) combinations in Table~\ref{tab:models} where we include convective 
overshoot, we provide nucleosynthesis predictions from 
the case with and without overshoot. If there is no TDU in the model 
without convective overshoot (e.g., 1.5$\Msun$, $Z = 0.014$) no \iso{13}C pocket is
included. For these ($M, Z$) combinations, the value of $N_{\rm ov}$  used in the 
calculation is provided in the header files. If no value of $N_{\rm ov}$ is specified,
no overshoot is included.

In the Appendix, we provide examples of each of the data file types.
Table~\ref{tab:examplesurf} illustrates the information included in the surface
abundance data files.  The surface abundance data tables start with the initial 
abundances used in the
post-processing calculations and then include elemental abundances as a function of
thermal pulse number.  At each entry we include the thermal pulse number, the
stellar mass, core mass and envelope mass at that thermal pulse (in $\Msun$), and 
the surface luminosity (in $\log \Lsun$). After the abundances of each element
are given, we then provide the surface elemental ratios of He/H, C/O and N/O at
that thermal pulse.  The final entry for each ($M, Z$) combination is the final
elemental abundances, computed at the last time step (which may fall on a thermal
pulse or during the interpulse period).

For all elements except Li, B and B we include the
element name, the proton number, $Z$; the abundance in the format 
$\log \epsilon (X)$ where $\log \epsilon (X) = \log_{10}(X/H) + 12$; [X/H]\footnote{where we use the
standard spectroscopic notation, [A/B] = $\log_{10}(A/B)_{\rm surf} - \log_{10}(A/B)_{\odot}$. 
The ratio $(A/B)_{\rm surf}$ is the number ratio of elements $A$ and $B$ at the surface of
the model star and $(A/B)_{\odot}$ is the solar number ratio, taken from \citet{asplund09}.}, 
[X/Fe], [X/O], and the mass fraction $X(i)$. The radioactive elements Tc and Pm may 
have non-zero $\log \epsilon (X)$ values, if they are produced
in the He-intershell and dredged to the surface. Note that we do not decay the abundances of 
radioactive isotopes (e.g., \iso{26}Al, \iso{60}Fe) in the isotopic or elemental surface 
abundance files, but they are assumed to have all decayed in the yield tables. 
However, we do decay the isotope \iso{93}Zr to \iso{93}Nb because Nb obtains essentially all 
of its production via this decay.

We do not include the surface abundances and yields for Be and B because these 
elements are not synthesized in stars. Theoretical predictions for Li are 
highly dependent on the numerical procedure, as demonstrated by \citet{lattanzio15} for
low-mass RGB stars with thermohaline mixing. The Li abundances in low-mass stars
is altered by thermohaline mixing or some other deep mixing process on the RGB, which
we do not include here. This means our models of low-mass AGB stars begin the AGB with
incorrect Li abundances implying that our AGB yields will similarly be incorrect. 
Note that while this problem also affects the \iso{12}C/\iso{13}C ratio, 
it is possible to correct for the effects of extra mixing on the evolution of the 
\iso{12}C/\iso{13}C ratio \citep[e.g.,][]{lebzelter08,karakas10b}. This is not possible for Li
because of the uncertainties affecting numerical predictions. Li abundance predictions from 
models with HBB have been shown to be consistent with observations of bright O-rich 
stars in the Magellanic Clouds and Galaxy \citep[e.g.,][]{lattanzio97,ventura00,vanraai12, garcia13}. 
This suggests that predictions are more robust in intermediate-mass models, where production 
can be copious. For this reason we provide lithium abundances and yields from models with HBB in
Table~\ref{tab:lithium}, with the results discussed in Section~\ref{sec:hetofe}.

In the isotopic data tables we provide the following isotopic ratios, many of which are useful
for comparison to stellar/PNe spectra or stardust grains:
\iso{12}C/\iso{13}C, \iso{14}N/\iso{15}N, \iso{16}O/\iso{17,18}O, 
\iso{24}Mg/\iso{25,26}Mg, \iso{26}Al/\iso{27}Al, \iso{28}Si/\iso{29,30}Si,
\iso{36,37}Cl/\iso{35}Cl, \iso{36,38}Ar/\iso{40}Ar, \iso{40,41}K/\iso{39}K, 
\iso{42,43,44,46,48}Ca/\iso{40}Ca, \iso{46,47,49,50}Ti/\iso{48}Ti,
\iso{53,54}Cr/\iso{52}Cr, \iso{54,57,58,60}Fe/\iso{56}Fe, and 
\iso{60,61,62}Ni/\iso{58}Ni.  In Table~\ref{tab:exampleiso} we show an example of the
isotopic data files available for each model, this time for the 3.5$\Msun$, $Z=0.03$
model. We include only 5 isotopic ratios for illustrative purposes over the first three
thermal pulses. The tables finish with  the final isotopic ratios calculated 
at the last time step.

\subsection{Elements from helium to iron} \label{sec:hetofe}

\begin{table*}
\begin{center}
\caption{Peak surface lithium abundance during the AGB and stellar yields from models 
with hot bottom burning. }
\label{tab:lithium}
  \vspace{1mm}
\begin{tabular}{cccccc} \hline \hline
Initial mass & $Z$ &  Peak lithium abundance & TP at peak & Mass Li expelled & Initial Li in wind \\
($\Msun$)    &     &  $\log \epsilon$~(Li)   &            & ($\Msun$) & ($\Msun$) \\
\hline \hline
4.75   & 0.03  & 3.00 & 15 & 1.7599538E-07 & 6.5591479E-08 \\ 
5.00   & 0.03  & 4.03 & 15 & 2.3315576E-07 & 6.9651541E-08 \\
5.50   & 0.03  & 4.70 & 14 & 1.4348146E-07 & 7.7758202E-08 \\
6.00   & 0.03  & 4.87 & 12 & 1.8557853E-07 & 8.5831154E-08 \\
7.00   & 0.03  & 5.18 &  5 & 2.3292797E-07 & 1.0173942E-07  \\
8.00   & 0.03  & 5.18 &  1 & 5.1344944E-07 & 1.1707634E-07  \\
\hline
4.25   & 0.014 &  3.80 & 19 & 4.1797335E-08 & 3.0904165E-08 \\
4.50   & 0.014 &  3.90 & 19 & 3.2354016E-08 & 3.3096789E-08 \\
4.75   & 0.014 &  3.95 & 18 & 2.8601756E-08 & 3.5298505E-08 \\
5.00   & 0.014 &  4.03 & 15 & 1.0902397E-08 & 3.7490221E-08  \\
5.50   & 0.014 &  4.17 & 13 & 1.7997854E-09 & 4.1848171E-08 \\
6.00   & 0.014 &  4.28 & 10 & 7.7072976E-10 & 4.6206125E-08  \\
7.00   & 0.014 &  4.40 & 5  & 9.1141705E-10 & 5.4802843E-08  \\
8.00   & 0.014 &  4.43 & 1  & 3.0000244E-08 &  6.3114797E-08 \\ 
\hline
4.00   & 0.007 &  2.91 & 18 & 9.1520405E-08 & 1.4334581E-08 \\
4.25   & 0.007 &  3.36 & 16 & 4.0745888E-08 & 1.5426464E-08  \\
4.50   & 0.007 &  4.01 & 15 & 8.2011226E-10 & 1.6500147E-08 \\
4.75   & 0.007 &  4.03 & 14 & 3.9599790E-10 & 1.7592029E-08 \\
5.00   & 0.007 &  4.06 & 13 & 1.9964203E-10 & 1.8674355E-08 \\
5.50   & 0.007 &  3.62 & 12 & 2.1940057E-11 & 2.0830822E-08 \\
6.00   & 0.007 &  4.42 &  7 & 4.0835935E-11 & 2.2946342E-08 \\
7.00   & 0.007 &  4.46 &  1 & 5.8903787E-10 & 2.7113233E-08  \\
7.50   & 0.007 &  4.20 &  1 & 2.2296360E-08 & 2.9042221E-08 \\
\hline \hline
\end{tabular}
\medskip\\
\end{center}
\end{table*}

In Table~\ref{tab:lithium} we present Li abundances and stellar
yields for models that experience HBB. We include the peak surface 
abundance of lithium, the thermal pulse number when the peak abundance 
occurs,  the mass of lithium expelled from the star (in $\Msun$), 
and the initial mass of lithium present in the wind (in $\Msun$). 
All of the models in Table~\ref{tab:lithium} experience HBB, as noted
in \citet{karakas14b}, with the exception of the 4.75$\Msun$, $Z = 0.03$
model. In \citet{karakas14b} the minimum temperature at the 
base of the envelope for HBB was arbitrarily set at 50~MK, which 
seems to be the minimum required to see the effects of CN cycling
at the surface. The peak temperature in the 4.75$\Msun$, $Z = 0.03$ 
model is 42.5~MK, which is hot enough to produce some Li with a peak 
$\log \epsilon$~(Li) = 3.0.   Table~\ref{tab:lithium}
shows that the peak Li abundance depends on the initial metallicity, with the
most massive, metal-rich $Z=0.03$ models predicting super-Li rich 
AGB stars with $\log \epsilon$~(Li) $\ge 5$.  The table
also shows that the peak abundance occurs earlier in the evolution
when the stellar mass increases, with models $M \ge 7\Msun$ experiencing the
highest abundance at the first thermal pulse.

\begin{figure}
    \begin{center}
          \includegraphics[width=0.95\columnwidth]{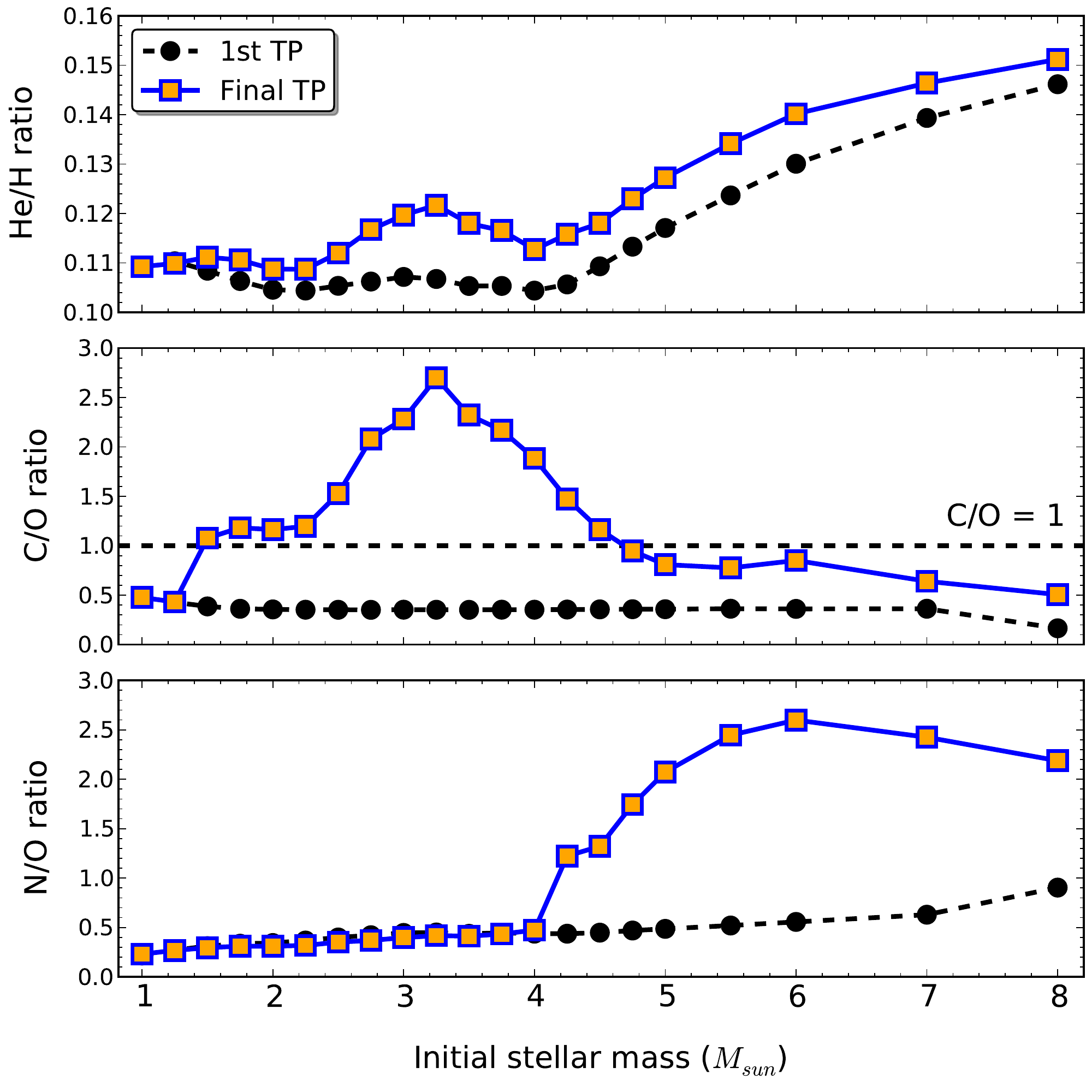}
\caption{The ratios of He/H, C/O and N/O at the surface after the first thermal 
pulse and after the last thermal pulse for the $Z=0.014$ models.}
    \label{z014-ratio1}
  \end{center}
\end{figure}

\begin{figure}
    \begin{center}
          \includegraphics[width=0.95\columnwidth]{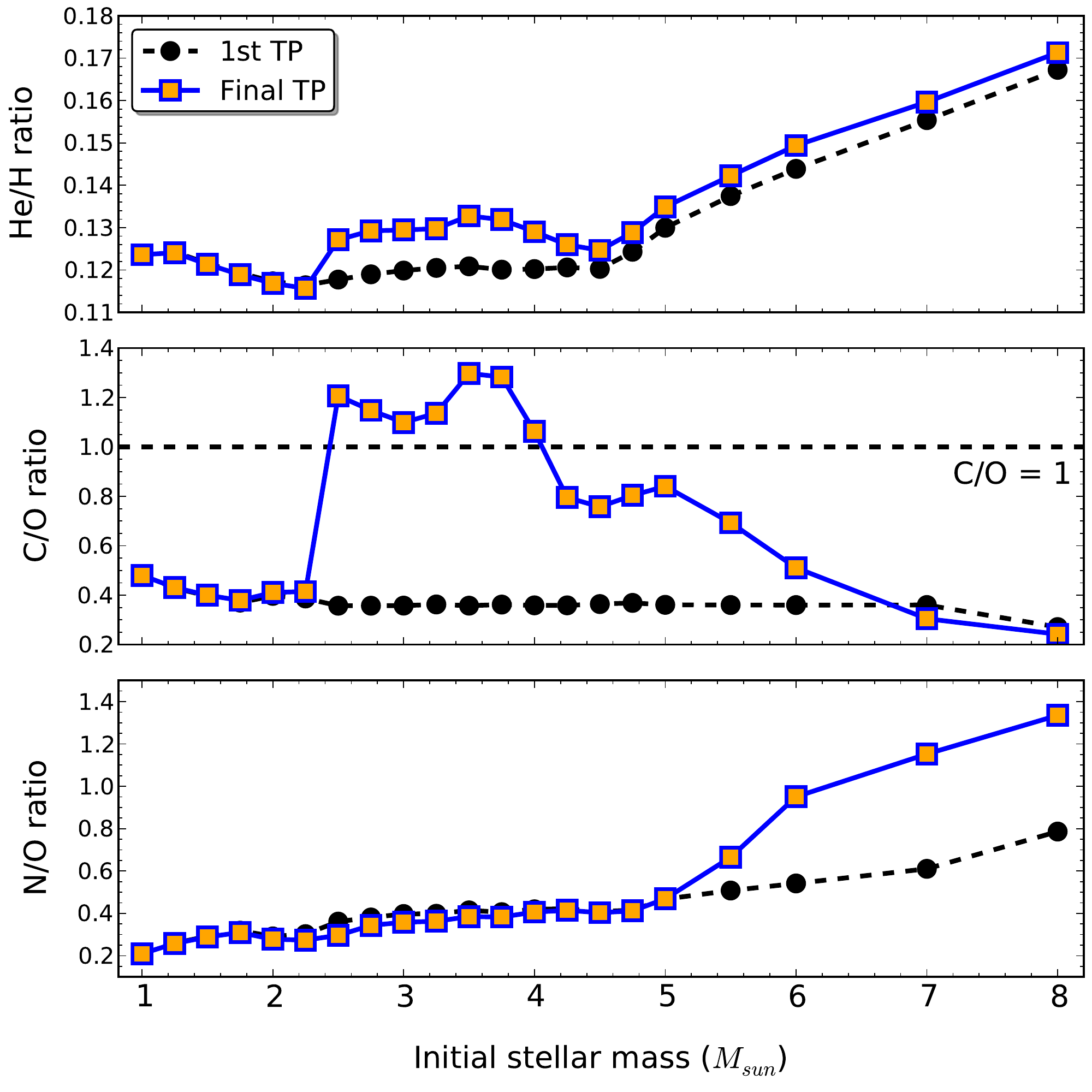}
\caption{Same as Fig.~\ref{z014-ratio1} except for the $Z=0.03$ models.}
    \label{z03-ratio1}
  \end{center}
\end{figure}

\begin{figure}
    \begin{center}
     \includegraphics[width=0.95\columnwidth]{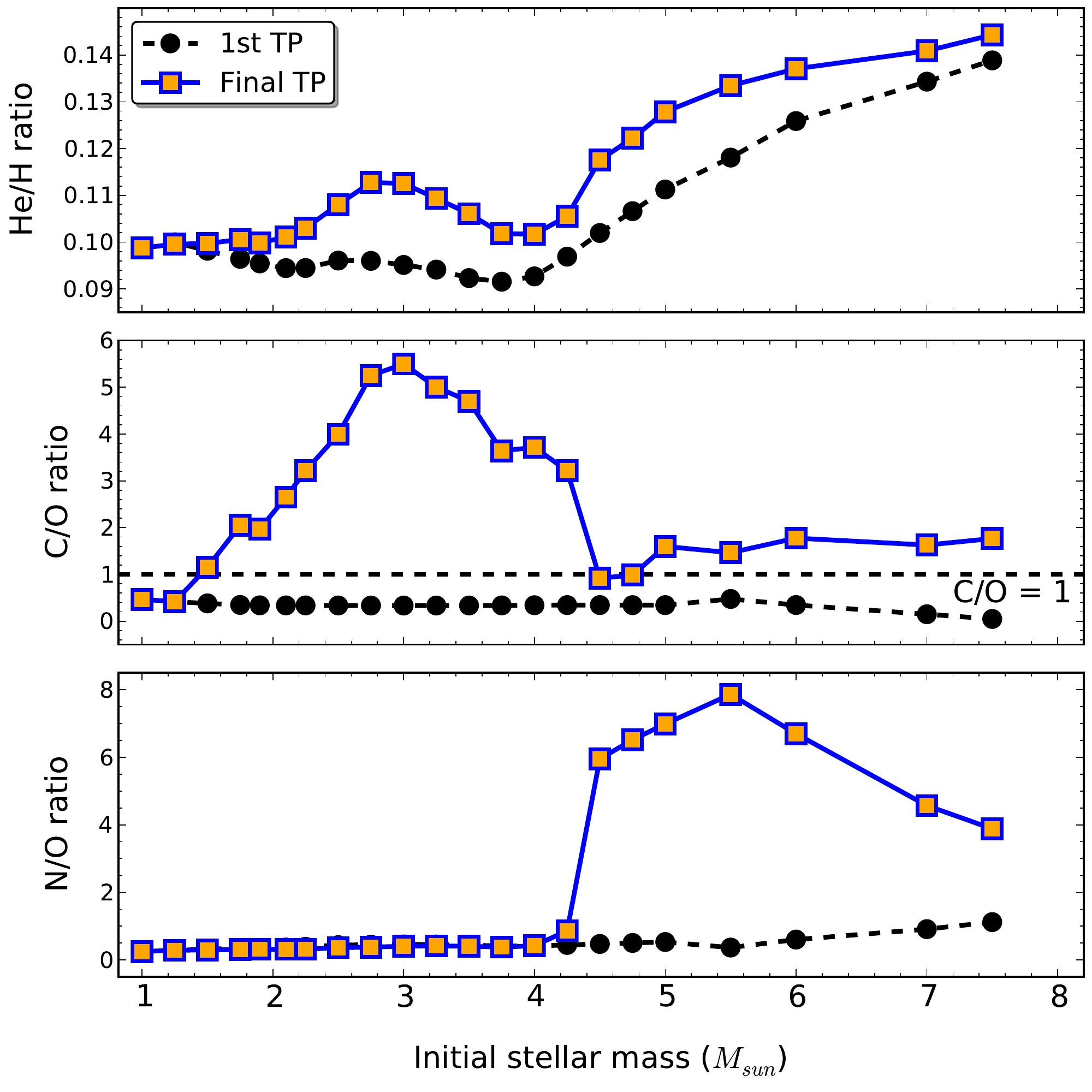}
\caption{Same as Fig.~\ref{z014-ratio1} except for the $Z=0.007$ models.}
    \label{z007-ratio1}
  \end{center}
\end{figure}

In Figs.~\ref{z014-ratio1},~\ref{z03-ratio1} and~\ref{z007-ratio1}  we show 
the He/H, C/O, and N/O ratios from the surface of the $Z=0.014$, $Z=0.03$ and $Z=0.007$ models. 
We include models with convective overshoot 
where applicable.  We plot the ratios after the first thermal pulse and after the last thermal pulse. 
The ratio after the first thermal pulse reflects surface abundances changes prior to the 
TP-AGB. This includes the first dredge-up (FDU), which occurs in all models although the surface abundance
changes are strongest around 2--3$\Msun$ \citep{boothroyd99} and the second dredge-up (SDU).
Second dredge-up occurs in models more massive than about $4-5\Msun$, depending on $Z$. 
We refer to Table~\ref{tab:models} and Table~1 from \citet{karakas14b} for the minimum
masses for SDU and HBB.  The 7 and 8$\Msun$ models begin HBB before the first thermal pulse, and 
this is reflected in the lower C/O and higher N/O ratios.  

\begin{figure}
    \begin{center}
          \includegraphics[width=0.95\columnwidth]{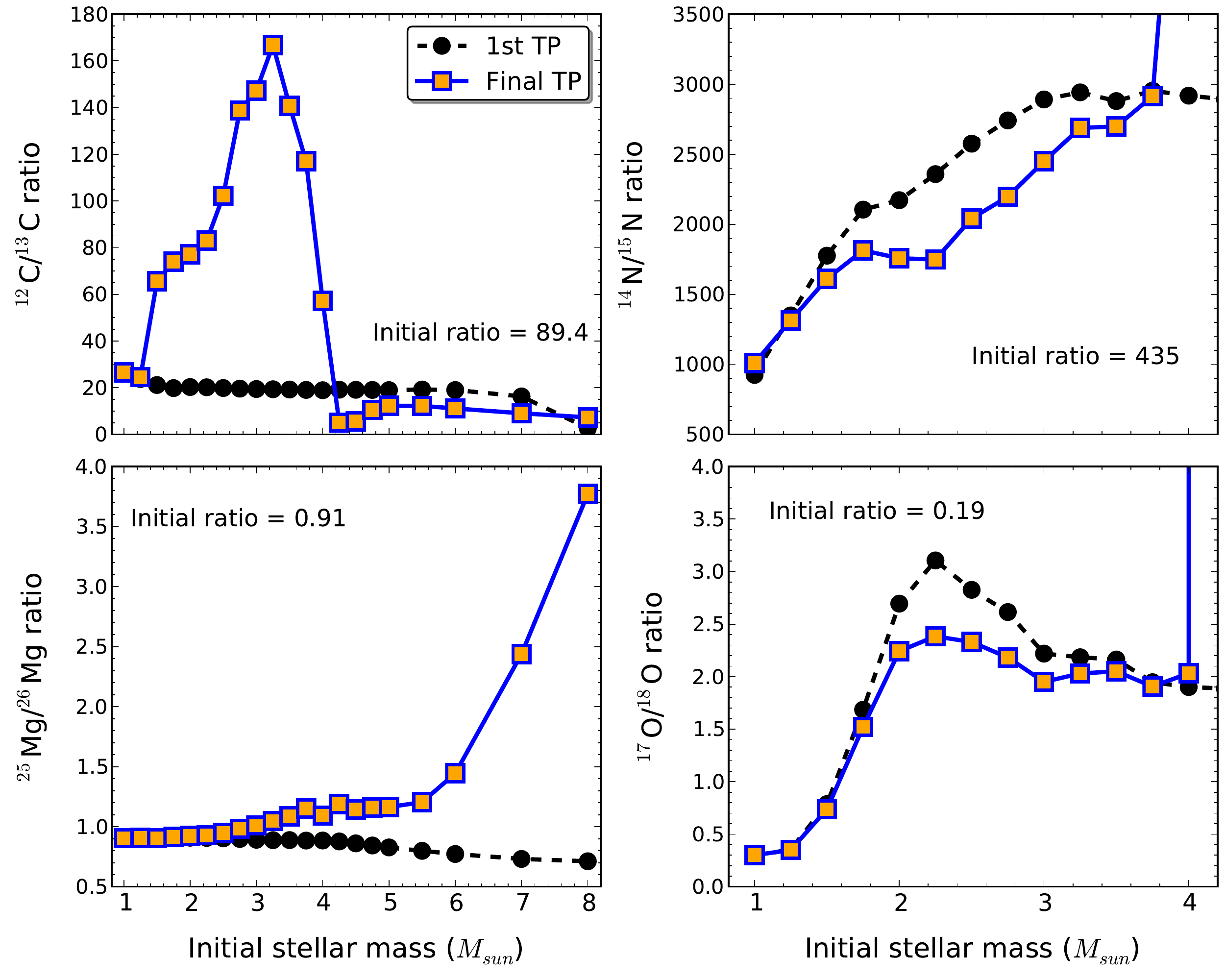}
\caption{The ratios of \iso{12}C/\iso{13}C, \iso{14}N/\iso{15}N, \iso{17}O/\iso{18}O,
and \iso{25}Mg/\iso{26}Mg at the first thermal pulse and after the
last thermal pulse for the $Z=0.014$ models. For the \iso{14}N/\iso{15}N and \iso{17}O/\iso{18}O
ratios, we show results for models without HBB, which includes masses up to 4$\Msun$.
The initial ratio is indicated on each panel.}
    \label{z014-ratio2}
  \end{center}
\end{figure}

\begin{figure}
    \begin{center}
    \includegraphics[width=0.95\columnwidth]{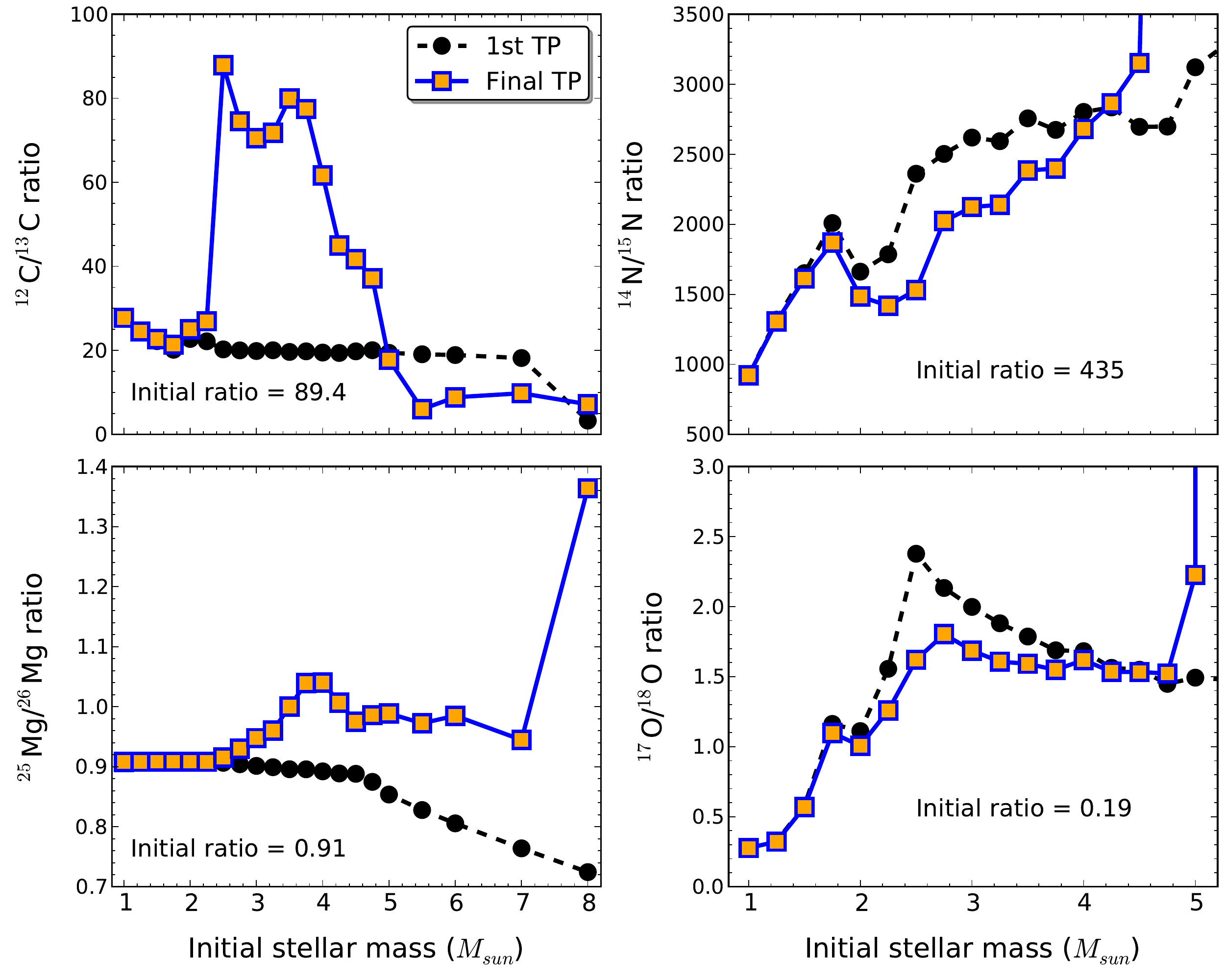}
\caption{Same as Fig.~\ref{z014-ratio2} except for the $Z=0.03$ models. 
For \iso{14}N/\iso{15}N and \iso{17}O/\iso{18}O,
we show the ratio for models without HBB, which includes masses up to 5$\Msun$.}
    \label{z03-ratio2}
  \end{center}
\end{figure}

\begin{figure}
    \begin{center}
     \includegraphics[width=0.95\columnwidth]{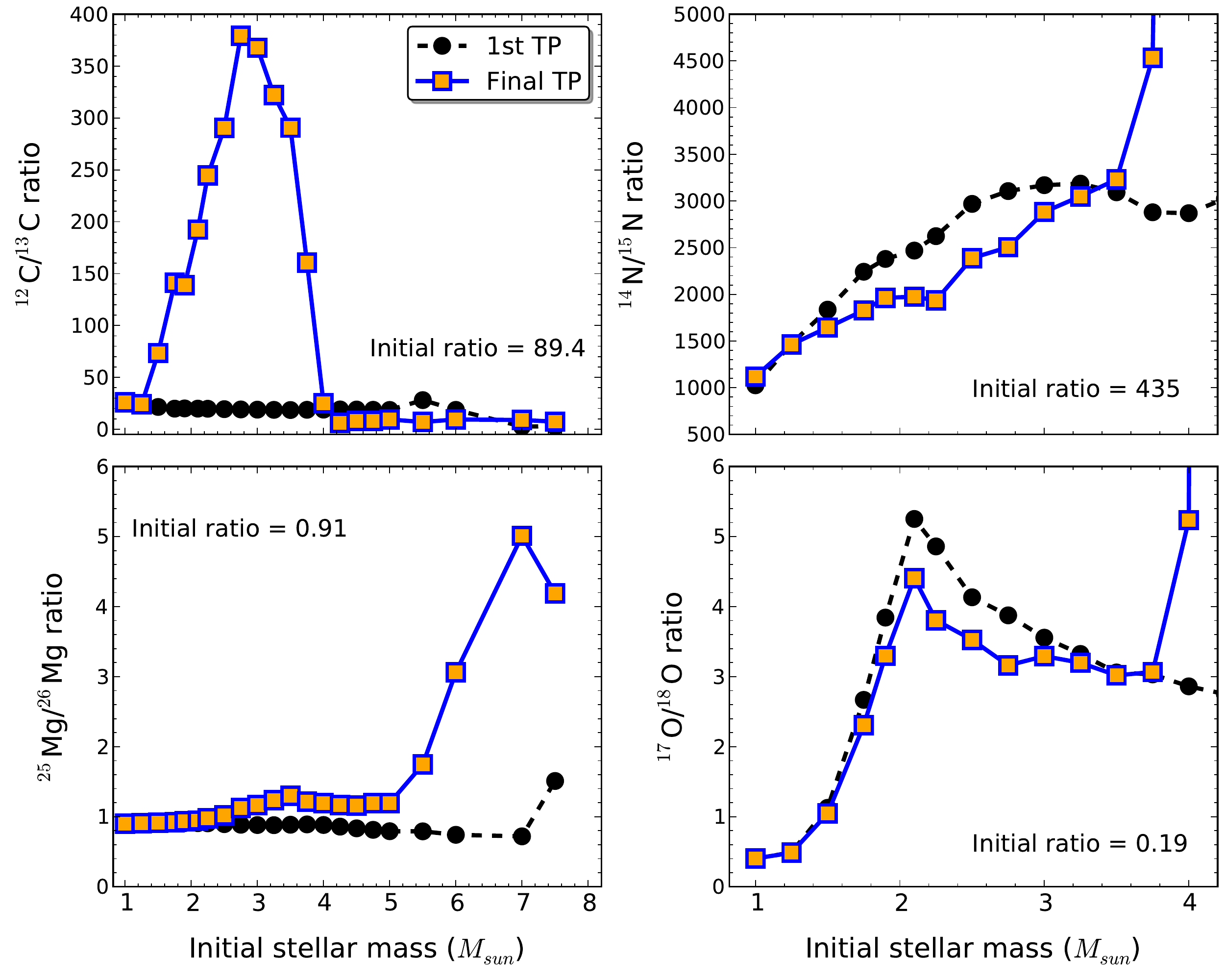}
\caption{Same as Fig.~\ref{z014-ratio2} except for the $Z=0.007$ models.}
    \label{z007-ratio2}
  \end{center}
\end{figure}

In Figs.~\ref{z014-ratio2},~\ref{z03-ratio2}, and~\ref{z007-ratio2} we show the ratios of 
\iso{12}C/\iso{13}C, \iso{14}N/\iso{15}N, \iso{17}O/\iso{18}O, and \iso{25}Mg/\iso{26}Mg at the 
first thermal pulse and after the last thermal pulse for the $Z=0.014$, $Z=0.03$, and $Z = 0.007$ 
models. We do not include any thermohaline mixing or other form of non-convective extra mixing into our 
calculations of RGB and AGB envelopes. This mean that our e.g., \iso{12}C/\iso{13}C ratios 
in low-mass ($M \lesssim 2\Msun$) models are higher than measured in RGB stars 
\citep[see discussions in][]{charbonnel94,charbonnel07a,eggleton08,karakas14dawes}. 
It is unclear how much extra mixing occurs in the envelopes of solar-metallicity AGB
stars that become carbon rich. Extra mixing on the RGB and TDU on the AGB can account
for the majority of the observed \iso{12}C/\iso{13}C ratios on the AGB \citep{karakas10b}. 
There are exceptions including a small sample of C-rich stars with
low \iso{12}C/\iso{13}C ratios $< 30$ and the J-star population \citep{abia97,lebzelter08}. 
For these objects, some form of extra mixing on the AGB is required 
although the mechanism responsible is not known 
\citep[and it is probably not thermohaline mixing, see][]{stancliffe10,busso10}. 
Extra mixing also on the AGB has been invoked to explain the composition
of roughly 10\% of meteoritic stardust oxide grains (the Group 2 grains), which show
depletions in \iso{18}O \citep{nollett03,palmerini11}.

\citet{halabi15} perform a comprehensive analysis of  model predictions against
observations of C, N and O isotopic ratios of Galactic RGB stars.  In their
study they found agreement between their solar-like composition models spanning a 
range from 1.2--7$\Msun$ and the predictions in \citet{karakas14dawes} for 
the \iso{16}O/\iso{17}O ratio after first and second dredge. The predictions illustrated
in Fig.~\ref{z014-ratio2} are specifically for the AGB phase but the abundances 
at the first thermal pulse are consistent with the post-FDU and SDU abundances from 
the $Z=0.02$ models from \citet{karakas14dawes}. 
The exceptions are for the 7$\Msun$ and 8$\Msun$ models because both
of these cases start to show the effects of HBB between the deepest extent of SDU and the 
first thermal pulse. The ratios affected include
\iso{12}C/\iso{13}C, \iso{16}O/\iso{18}O and \iso{14}N/\iso{15}N. For example at the deepest
extent of SDU, the \iso{12}C/\iso{13}C = 19.3 at the surface of the 8$\Msun$, $Z = 0.014$ 
model; this drops to 2.64 by the start of the thermally-pulsing phase.

As discussed by \citet{halabi15} there are few observational constraints on the 
\iso{14}N/\iso{15}N ratio, especially for RGB stars. \citet{hedrosa13} measured the
\iso{14}N/\iso{15}N ratio in a sample of AGB stars and found evidence that some \iso{15}N
production occurs in C-rich AGB stars. Figs.~\ref{z014-ratio2},~\ref{z03-ratio2} 
and~\ref{z007-ratio2} show that the \iso{14}N/\iso{15}N ratio decreases between the
first thermal pulse and the tip of the AGB for models that become C-rich. This 
indicates that some \iso{15}N production is happening in our models, albeit at
a lower level than needed by the observations. The minimum value reached in our models is 1400,
while some stars have ratios $< 1000$.
The complex reaction pathway that produces \iso{19}F \citep{lugaro04} first produces \iso{15}N as 
an intermediate step. 
In models where the nitrogen
isotopic ratio decreases during the AGB, some of the \iso{15}N survives He-shell burning. In
higher mass models, the \iso{15}N is destroyed to make \iso{19}F (which itself may
be destroyed by $\alpha$ capture), or HBB destroys \iso{15}N by proton capture in the envelope.
Uncertainties in the reaction rates involved in this path may play a role, 
specifically the \iso{15}N($\alpha$,$\gamma$)\iso{19}F reaction.

The oxygen isotope ratios have been measured in RGB and AGB stars. The most comprehensive
study was by Harris and collaborators using high-resolution near-IR spectra 
\citep{harris84, harris85a, harris85b, harris87} for a sample of G, K, and M giants.
\citet{lebzelter15} used near-IR spectra to measure oxygen in a small sample
of RGB cluster stars covering a range of mass from 1.8--4.5$\Msun$. They found 
good agreement with the theoretical models of \citet{cristallo11} and the observed 
\iso{16}O/\iso{17}O ratio, but not with \iso{16}O/\iso{18}O. 
Many studies have focused on measuring the \iso{17}O/\iso{18}O ratio from CO 
observations in mass-losing AGB stars \citep{kahane92,kahane00,decin10,khouri14}, 
which is why we show show this ratio in Figs.~\ref{z014-ratio2},~\ref{z03-ratio2} 
and~\ref{z007-ratio2}.   \citet{kahane92} find values between
1.12 and 1.66 for the \iso{17}O/\iso{18}O ratio, which suggests stars with masses 
between $\approx 1.5\Msun$ to 2$\Msun$ at solar metallicity (or between 2--3$\Msun$ if
the stars have a metallicity that is $0.03 \approx 2Z_{\odot}$). The \iso{17}O/\iso{18}O
ratios estimated from the data of \citet{lebzelter15}
range from 1.08 for Star 415 in NGC~7789, with an
estimated mass of 1.8$\Msun$, to 1.42 for HD16068 in Tr2, which has an estimated mass 
of 4.5$\Msun$.  The observed ratios are lower than the predictions in 
Figs.~\ref{z014-ratio2} for their estimated mass, 
indicating that higher initial \iso{18}O abundance may be required, as also 
discussed by \citet{lebzelter15}.

Similar to the case for the nitrogen isotopic ratio, the model predictions 
illustrated in Figs.~\ref{z014-ratio2},~\ref{z03-ratio2} 
and~\ref{z007-ratio2} show that the  \iso{17}O/\iso{18}O ratio decreases
from the first thermal pulse to the tip of the AGB for models that become C-rich and
do not experience HBB. The decrease occurs because \iso{18}O is not completely 
destroyed by \iso{18}O($\alpha$,$\gamma$)\iso{22}Ne during TPs and is therefore
dredged to the stellar surface by TDU.
For models with HBB, efficient destruction of \iso{18}O leads
to very high predicted ratios. Measurements of the oxygen isotope ratio in bright O-rich
AGB stars has been attempted \citep{justtanont15}, with the result of a non-detection 
of \iso{18}O as evidence for the existence of HBB, which is consistent with calculations
\citep[see also][]{justtanont13}. 

Solar metallicity models near the minimum mass for HBB ($M \lesssim 4.75\Msun$) show final surface 
abundances that are consistent with J-type AGB stars and 
also a fraction of stardust silicon carbide (SiC) grains (the A$+$B grains), where C/O $\gtrsim 
1$ and \iso{12}C/\iso{13}C $\le 10$ \citep{amari01c}. These conditions are not met in models of 
higher metallicity (e.g., see Figs.~\ref{z03-ratio1} and~\ref{z03-ratio2} for the $Z=0.03$ models). 
The final surface composition of the intermediate-mass $Z=0.007$ models also have C/O $> 1$ and low 
\iso{12}C/\iso{13}C ratios, but the measured silicon isotopic ratios of A$+$B grains suggest that 
they are mostly made in solar-metallicity stars \citep{amari01c}. Note however that models with HBB 
and C/O $\gtrsim 1$ also show high nitrogen and oxygen isotopic ratios. The nitrogen isotopic ratio 
in A$+$B grains covers orders of magnitudes, from $\sim$ 30 to $\sim$ 12,000 and the grains with the 
highest ratios may be compatible with these intermediate-mass models. This ratio is difficult 
to measure in real stars and would be especially challenging in bright AGB stars with HBB. 
We predict that \iso{15}N would not be detected, making a determination of the nitrogen 
isotopic ratio impossible.

The final \iso{25}Mg/\iso{26}Mg ratio is greater than solar in all the models
and strongly increases with the initial
mass as illustrated in, e.g., Fig.~\ref{z014-ratio2} for the solar metallicity models. 
In the models without HBB it is dominated by the production of these isotopes in the 
intershell via \iso{22}Ne+$\alpha$ reactions \citep{karakas06a}. 
In models with HBB the \iso{25}Mg/\iso{26}Mg ratio is also affected by proton captures 
and the operation of the MgAl chain. In both cases, production of \iso{25}Mg is favoured
consistent with the models of \citet{ventura13}.

\begin{figure}
    \begin{center}
     \includegraphics[width=0.95\columnwidth]{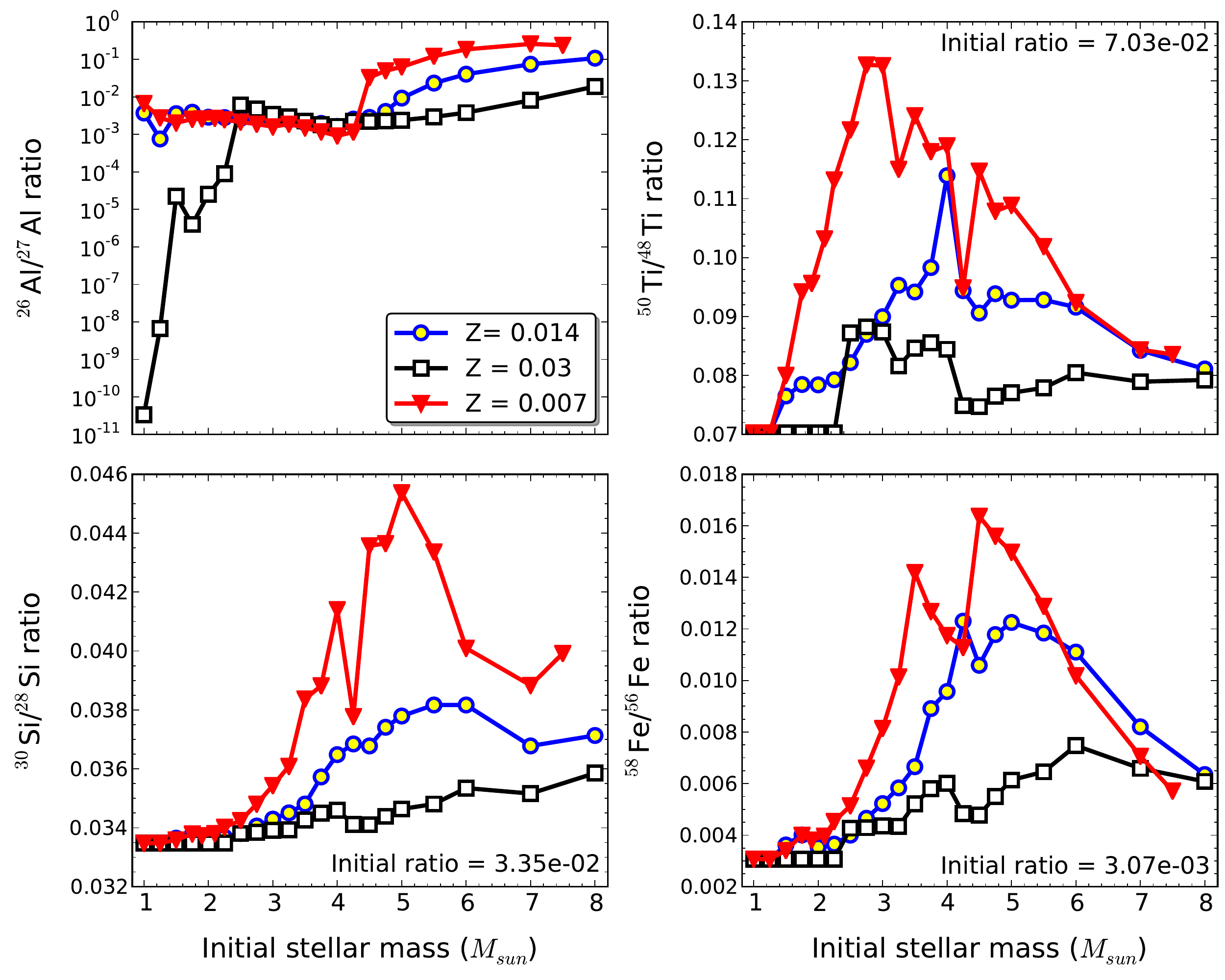}
\caption{Four selected examples of surface isotopic ratios after the final TP as function of 
the initial stellar mass for all three metallicities.} 
    \label{fig:lightiso}
  \end{center}
\end{figure}

Depending on the initial stellar mass several isotopic ratios of the elements from Al to Ni 
are predicted to show large variations. In Fig.~\ref{fig:lightiso} we present 
four examples of these isotopic ratios which are modified by AGB 
nucleosynthesis.  These can be measured in meteoritic stardust grains \citep[e.g.][]{amari01a},
Furthermore, the Si isotopic ratios can be measured in AGB stars via molecular lines 
\citep[e.g., from SiO,][]{peng13}. 
The typical value of the $^{26}$Al/$^{26}$Al ratio in AGB stars that do not experience HBB 
is between 10$^{-3}$ and 10$^{-2}$. This generally matches the range covered
by oxide grains from O-rich AGB stars and by SiC grains from C-rich AGB stars, although
it is not clear how to obtain values as low as 10$^{-4}$, which have been detected in 
some grains \citep{vanraai08a}.  

Most of the $^{26}$Al at the stellar
surface in these low-mass AGB models is the result of the TDU, where $^{26}$Al in the intershell
comes from the ingestion of the H-burning ashes. In the TP, $^{26}$Al can be
destroyed by neutron captures if the $^{22}$Ne neutron source is activated 
because the $^{26}$Al(n,p)$^{26}$Mg and $^{26}$Al(n,$\alpha$)$^{23}$Na reactions have
relatively high cross-sections ($\sim$ 200 mbarn).   The slight
decrease of $^{26}$Al/$^{26}$Al with stellar mass and the difference between the $Z=0.03$ models
and the other metallicities is consistent with the fact that 
the temperature at the base of the TP and
the activation of the $^{22}$Ne source increases with mass and decreases with metallicity.
For models with $Z=0.03$ and mass below 2.5 $\Msun$, the $^{26}$Al/$^{26}$Al ratio is much lower
because they do not experience efficient TDU. Because the initial abundance of 
$^{26}$Al is zero, it is extremely sensitive to the exact depth of the penetration of
the convective envelope during the TDU, even just into the tip of the thin region of  
H-burning ashes that is not mixed into the TP. This explains why
the 1 and 1.25 $\Msun$ models of $Z=0.007$ and $Z=0.014$ show higher $^{26}$Al/$^{27}$Al ratios than the 
models of the same mass but $Z=0.03$, and why the Z=0.03 models of mass
below 2.5 $\Msun$ show some variations with increasing the stellar mass. Above roughly 
4 $\Msun$, HBB is the dominant production channel for $^{26}$Al, 
and the $^{26}$Al/$^{27}$Al ratio reaches above 0.1. The $^{26}$Al/$^{27}$Al ratio grows 
with the temperature at the base of the envelope, which increases with increasing stellar mass and 
decreasing metallicity.

In contrast to \iso{26}Al/\iso{27}Al, the other three isotopic ratios shown in Fig.~\ref{fig:lightiso} are 
only affected by neutron captures. This is generally the case for the isotopic ratios of the elements 
below Fe in AGB stars, particularly at the metallicities discussed here. The $^{30}$Si/$^{28}$Si 
and the $^{58}$Fe/$^{56}$Fe ratios are mainly affected by the neutrons released in the TPs by the 
$^{22}$Ne neutron source, and also this is generally the case for most of the isotopic ratios of 
the elements below Fe in AGB stars. This results in the largest changes observed at the higher 
masses, with a peak around 5 to 6 $\Msun$, depending on the metallicity. The maximum variations 
for the $^{30}$Si/$^{28}$Si ratio is only 40\%, while for the $^{58}$Fe/$^{56}$Fe ratio it 
is a factor of 5. This is because the neutron-capture cross-sections in the region of Si are 
smaller than those in the region of Fe, and because the initial $^{58}$Fe abundance is very low. 
That the $^{30}$Si/$^{28}$Si ratio increases with a decrease in the metallicity is one piece of
evidence for an origin of SiC grains of type Y and Z in AGB stars of metallicity lower than solar 
\citep{hoppe97,amari01a}.

Finally, we show the peculiar case of the $^{50}$Ti/$^{48}$Ti ratio. Because
$^{50}$Ti is neutron magic ($N=28$), its neutron-capture cross-section is more than a factor of ten
smaller than those of the other Ti isotopes. This makes the $^{50}$Ti/$^{48}$Ti ratio a unique case 
among the isotopic ratios below Ni, in being sensitive to the neutron flux in the $^{13}$C pocket. 
Enhancements in this ratio can reach up to a factor of two and the maximum corresponds to 
models of mass 3--4 $\Msun$  because in this mass range the 
$^{13}$C neutron source is active.

In the top panel of Figs.~\ref{z014-surf},~\ref{z03-surf} and~\ref{z007-surf} we show the final 
surface composition 
for elements lighter than Fe for a selection of stellar  evolutionary sequences.
The figure illustrates that low-mass stars with $M \le 3\Msun$ produce substantial C, N, and F 
and some Ne and Na, where production increases with decreasing metallicity 
\citep[e.g.,][]{karakas10a,cristallo11}.   This is easily 
understood by examination of Fig.~\ref{mdredge} which shows that the lowest metallicity 
$Z=0.007$ dredge-up more intershell material at a given mass. The H and He-intershells of 
these models are also hotter, owing to a lower opacity. The intermediate-mass stars ($M > 3\Msun$)
show substantial N production and varying degrees of C, O and F destruction. The lowest
metallicity $Z=0.007$ intermediate-mass stars also show a small production of Al, where it is 
well known that the Mg-Al chain is more effective at lower metallicity owing 
to hotter HBB temperatures \citep[e.g.,][]{ventura11b}.
For the intermediate-mass elements between Si and Fe there are almost no changes in the 
elemental abundance as a 
consequence of AGB nucleosynthesis \citep[e.g.,][]{cristallo09,cristallo11,karakas09,karakas12,shingles13}, 
although there is a small production of P and Sc (e.g., at the level of [Sc/Fe] $\lesssim 0.3$) as a 
result of neutron captures \citep{smith89}.

\subsection{Elements heavier than iron}

\begin{figure}
    \begin{center}
    \includegraphics[width=0.95\columnwidth]{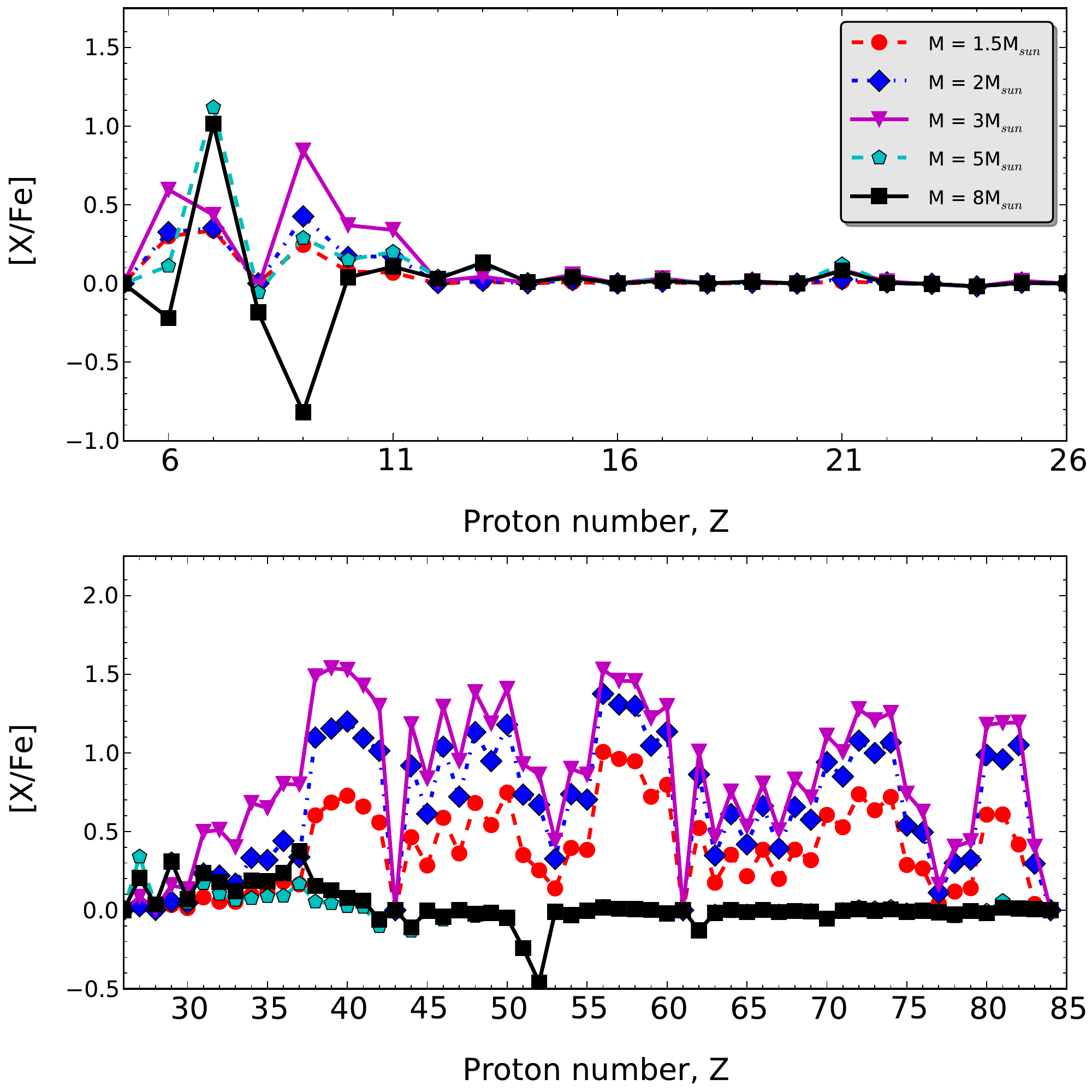}
\caption{Final elemental surface abundance (in [X/Fe]) for a selection of evolutionary models
with a solar metallicity, $Z=0.014$. The masses shown in the figure are indicated in the legend.
The top panel shows elements up to the iron peak and the lower panel elements heavier than iron.}    
\label{z014-surf}
  \end{center}
\end{figure}

\begin{figure}
    \begin{center}
   \includegraphics[width=0.95\columnwidth]{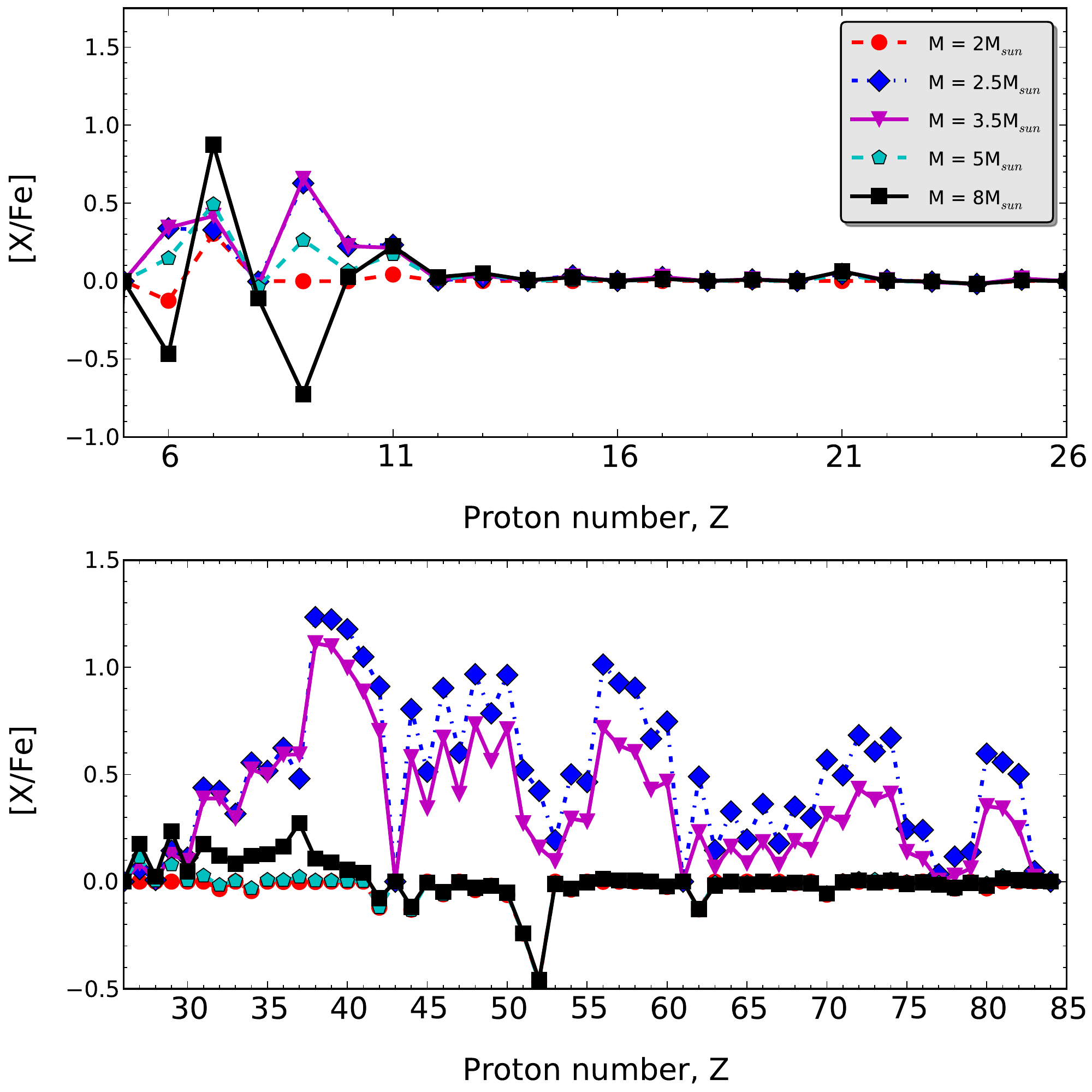}
\caption{Same as Fig.~\ref{z014-surf} for a selection of $Z=0.03$ models.}
    \label{z03-surf}
  \end{center}
\end{figure}

\begin{figure}
    \begin{center}
   \includegraphics[width=0.95\columnwidth]{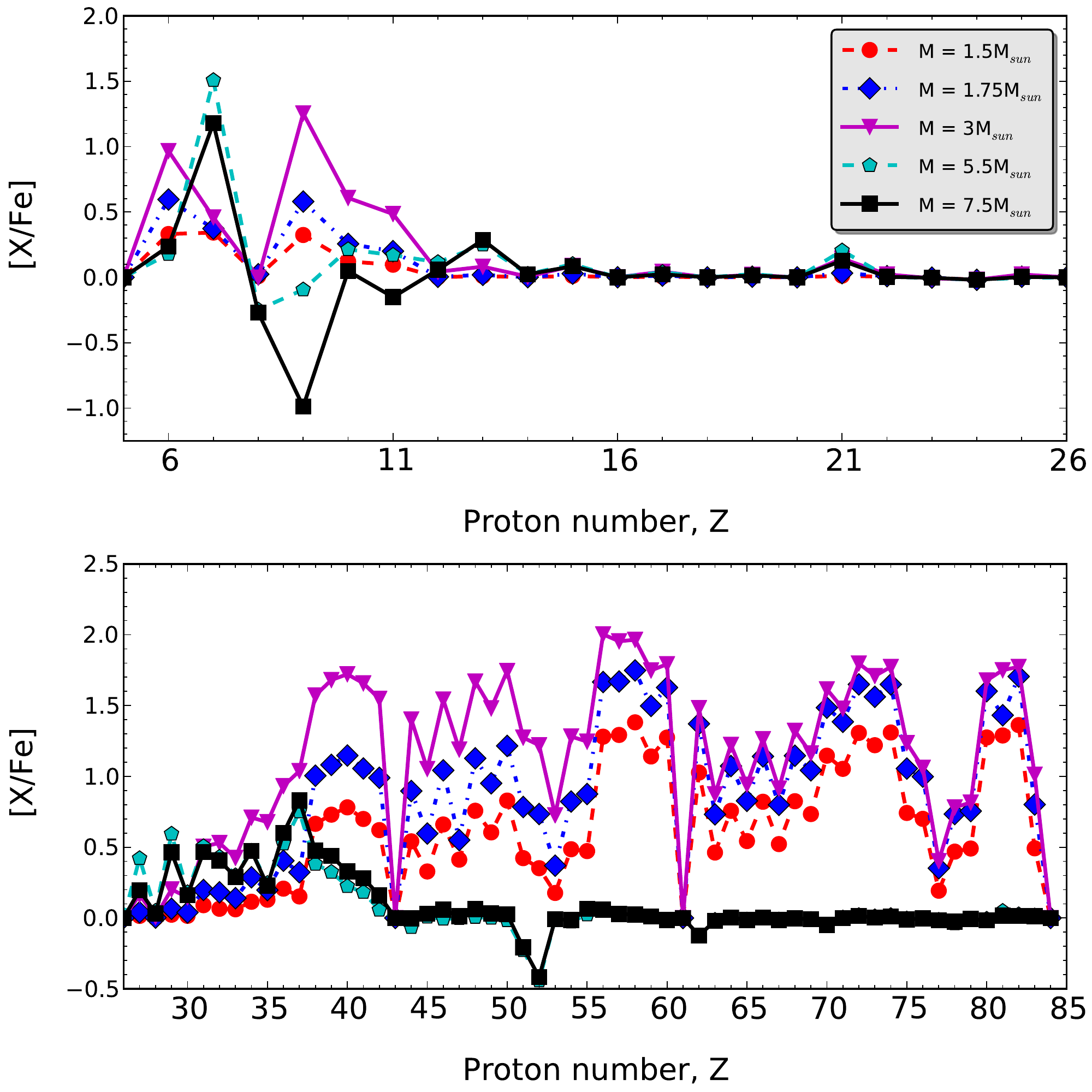}
\caption{Same as Fig.~\ref{z014-surf} for a selection of $Z=0.007$ models.}
    \label{z007-surf}
  \end{center}
\end{figure}

In the lower panel of Figs.~\ref{z014-surf},~\ref{z03-surf} and~\ref{z007-surf} we show the final surface 
composition for elements heavier than Fe. For the metallicities in our study, intermediate-mass
models with masses above 4.5$\Msun$ without 
\iso{13}C pockets show little production of $s$-process elements. 
This is not the case for lower metallicities $Z \le 0.001$ \citep{lugaro12,fishlock14b,shingles15}.
Some production around the first $s$-process peak at Rb (number of protons $z=37$) 
occurs, although it is generally [Rb/Fe] $\lesssim 0.5$ for all solar metallicity models, 
even when a \iso{13}C pocket is included.  The intermediate-mass models of $Z=0.014$
predict lower Rb enrichments than models of the same mass and similar metallicity 
($Z=0.02$, [Fe/H] = $+0.14$) from \citet{karakas12}, except for the 5$\Msun$ case. 
This is because here we are using an updated \iso{22}Ne($\alpha$,n)\iso{25}Mg reaction 
from \citet{iliadis10}, whereas in \citet{karakas12} we were using the faster 
NACRE rate \citep{angulo99}. Some production of the elements between Fe and Sr occurs, 
which is typically associated with the weak $s$-process in massive stars, e.g., Ga ($z=31$) 
and Ge ($z=32$), as well as peaks at Co ($z=27$) and Cu ($z=29$).

In the low-mass models that include \iso{13}C pockets
the $s$-process production and distribution is strongly dependent on the initial stellar metallicity, as
discussed in \citet{busso01}. Here we focus on metallicities near solar, where production is
dominated by elements at the first $s$-process peak, Sr, Y, and Zr ($z=38, 39, 40$), 
and the second peak, at Ba, La, and Ce ($z=56, 57, 58$). 
This is especially the case for solar and super-solar
metallicities as illustrated in Figs.~\ref{z014-surf} and~\ref{z03-surf}. In the 
lower metallicity $Z=0.007$ we see a noticeable shift in the predicted distribution of $s$-process 
elements with more 2nd-peak elements made and a stronger production of Pb ($z=82$).  

The shift in the $s$-process distribution can be quantified by calculating the $s$-process relative 
indicators: [hs/ls] and [Pb/hs], where ``ls'' stands for light $s$-process elements
(Y, Sr, Zr) and ``hs'' for heavy $s$-process elements (Ba, La, Ce). The
[Rb/Zr] ratio provides further information on the neutron density in the He-shell. We do not provide
[hs/Fe] and [hs/Fe] in the surface abundance data files, owing to the fact that there are
a number of different choices of the elements included in the definition 
in the literature \citep{bisterzo10,cristallo11,lugaro12}.
We leave it to readers to calculate these indicators as they choose. However, in the yield tables we 
provide the [Rb/Zr], [ls/Fe], [hs/Fe], [hs/ls], and [Pb/ls] ratios, calculated from the
integrated yields for each ($M, Z$) combination, using the elements listed above \citep{lugaro12}.

We come back to demonstrating how these ratios vary with mass and metallicity 
in Section~\ref{sec:yields}, here we discuss an illustrative example using the 3$\Msun$ models.  
The advantage of using the $s$-process relative indicators is that they are largely free of the
uncertainties related to the stellar modelling, such as the mass-loss rate and the TDU efficiency 
\citep{lugaro12}. This is important because we see from Fig.~\ref{mdredge} that the amount of 
material dredged to the surface increases as a function of metallicity for models around 3$\Msun$.
The [hs/ls] ratios are $-0.238$ from the $Z=0.03$ model, $-0.026$ from the
solar metallicity model, and $+0.320$ from the $Z=0.007$ model. This shows the shift from an $s$-process 
distribution dominated by the first peak in super-solar metallicities stars of [Fe/H] = $+0.3$, 
to a distribution dominated by the second peak in stars of [Fe/H] = $-0.3$. 
The distribution in the solar metallicity stars is fairly balanced
between the first and second peaks. Furthermore, the [Pb/hs] ratio increases
from $-0.389$, $-0.279$ and $-0.198$ in the $Z=0.03, 0.014$, and $Z=0.007$ models respectively,
showing an increase in the Pb production relative to the second peak.
 
\section{Stellar Yields} \label{sec:yields}

\begin{figure}
    \begin{center}
          \includegraphics[width=0.95\columnwidth]{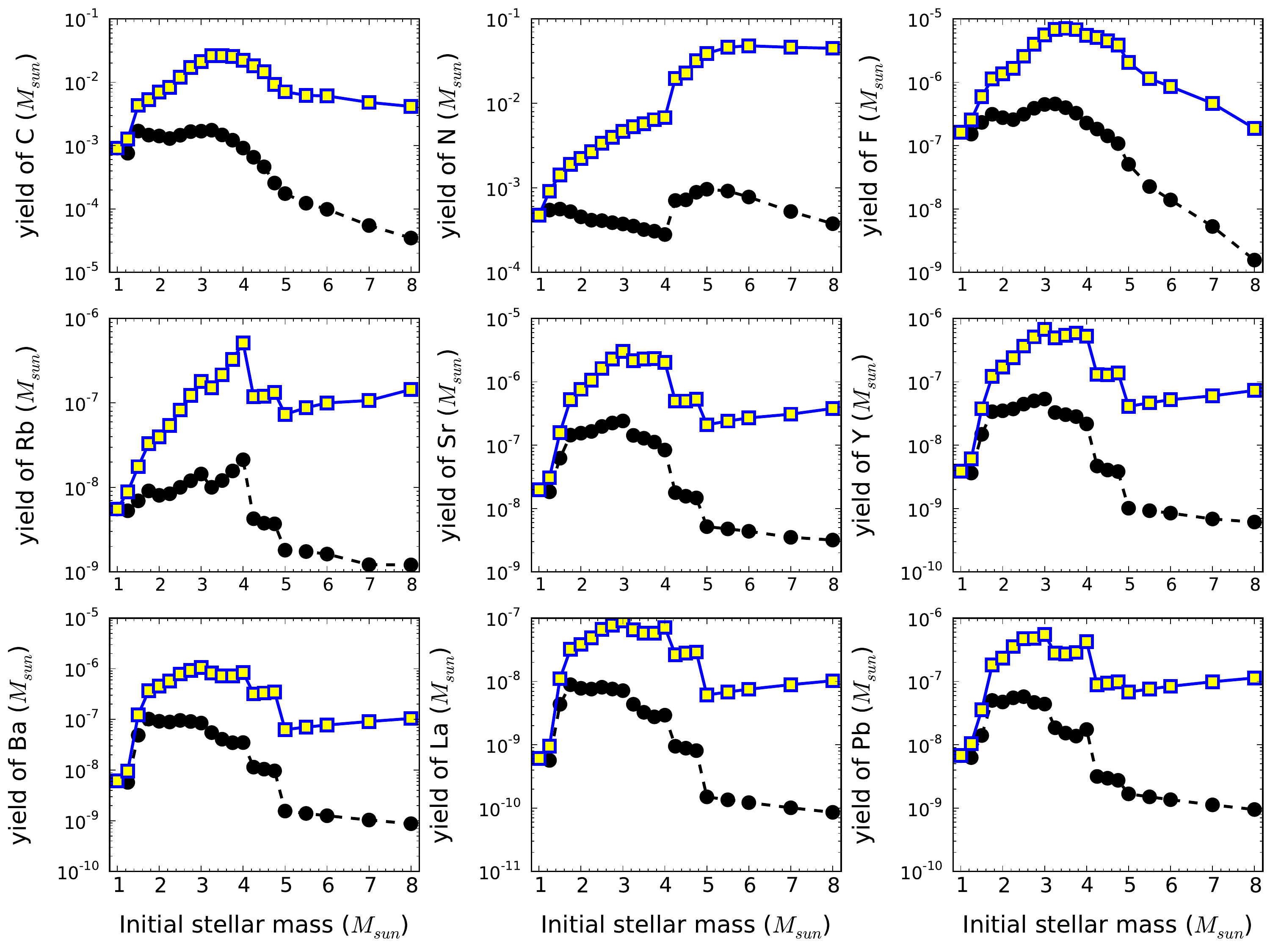}
\caption{Stellar yields from the $Z=0.014$ models shown as a function of the initial stellar mass. 
The yield is the total mass expelled (in $\Msun$) for a selection of elements (blue-yellow squares) 
compared to the yields weighted by the Salpeter IMF (black filled circles).}
    \label{z014-yield}
  \end{center}
\end{figure}

\begin{figure}
    \begin{center}
          \includegraphics[width=0.95\columnwidth]{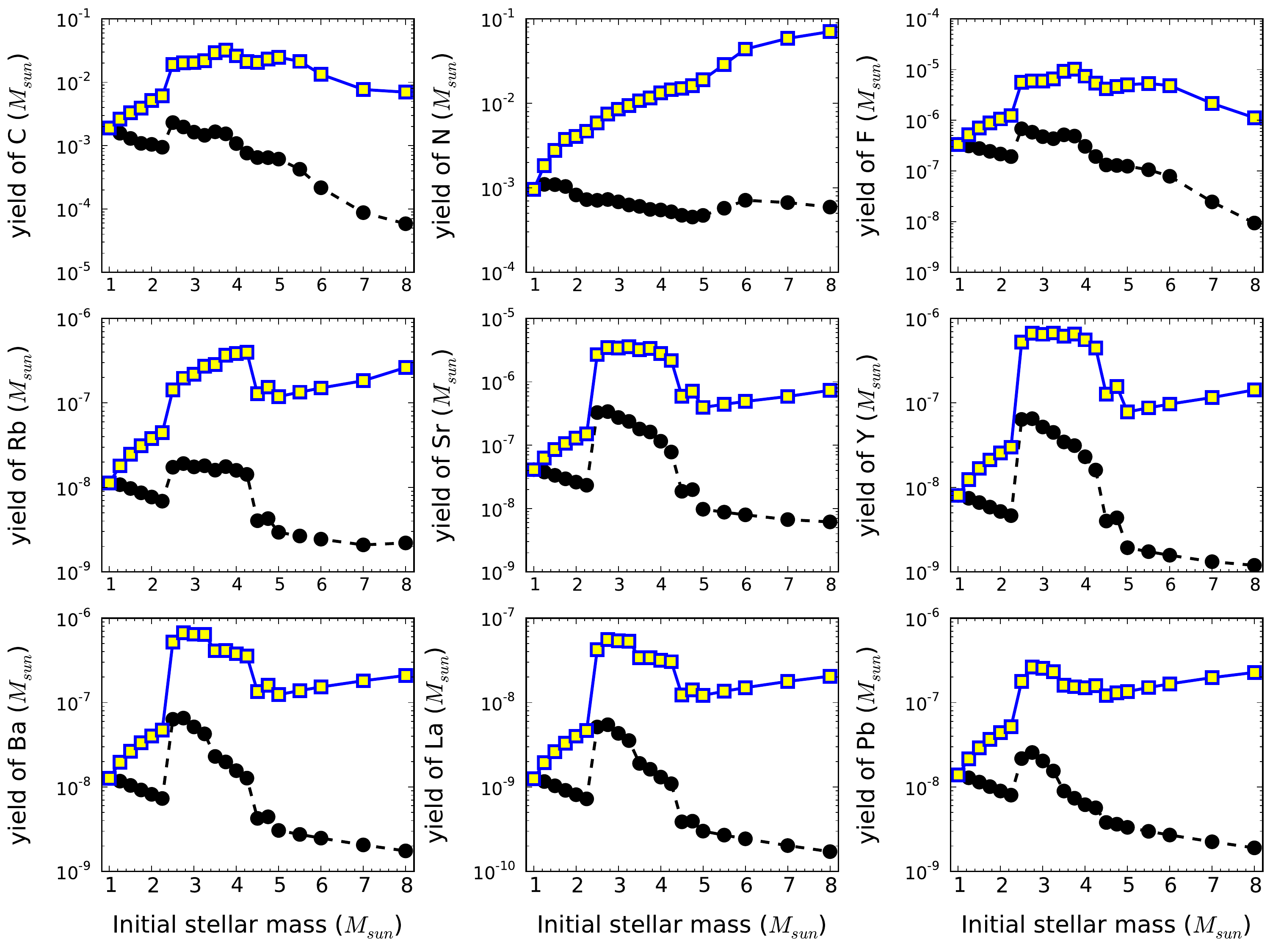}
\caption{Same as Fig.~\ref{z014-yield} but for the $Z=0.03$ models.}
    \label{z03-yield}
  \end{center}
\end{figure}

\begin{figure}
    \begin{center}
          \includegraphics[width=0.95\columnwidth]{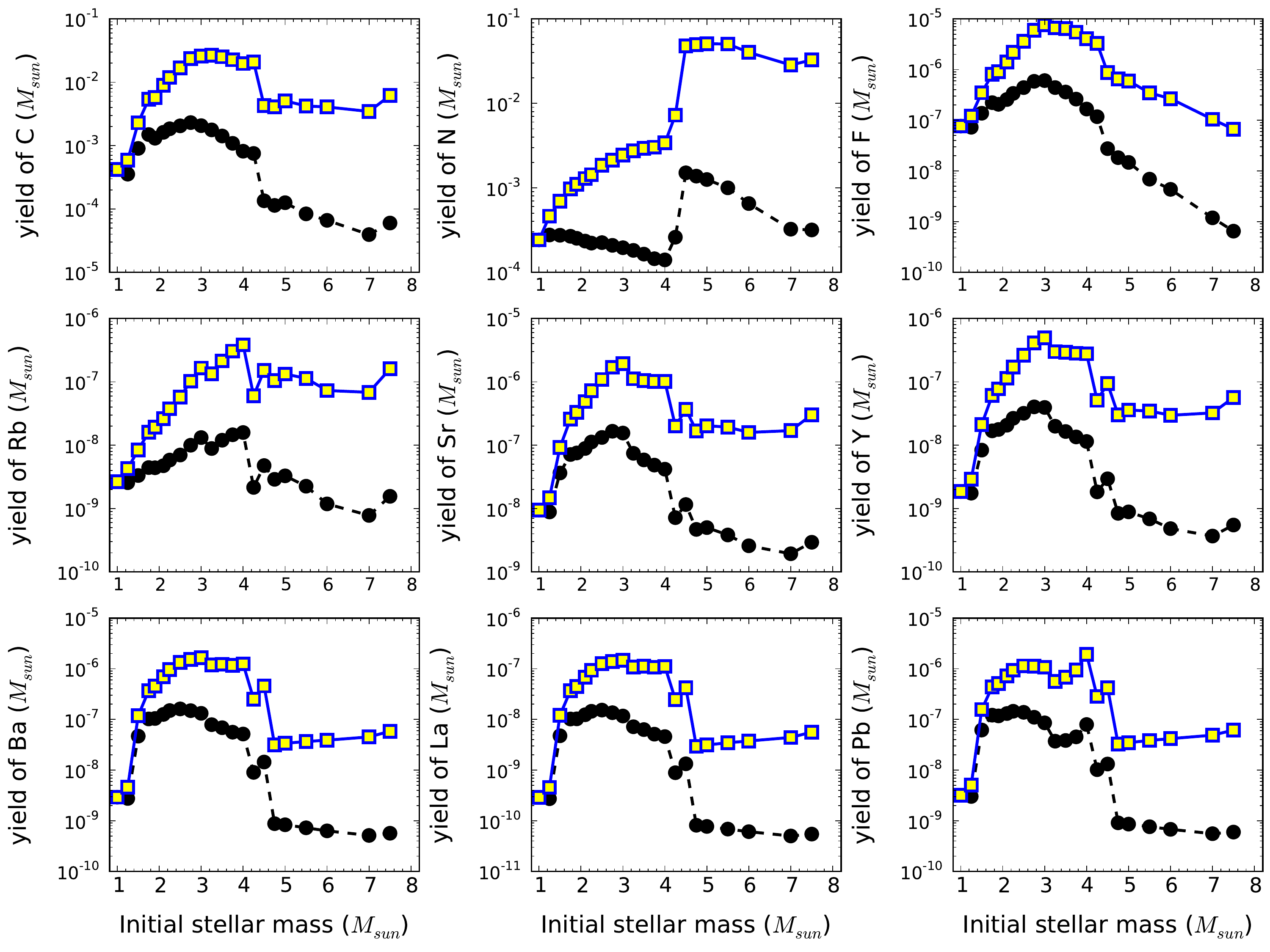}
\caption{Same as Fig.~\ref{z014-yield} but for the $Z=0.007$ models.}
    \label{z007-yield}
  \end{center}
\end{figure}

Stellar yields are an essential ingredient for theoretical calculations of Galactic chemical
evolution \citep{tinsley80,romano10,nomoto13,karakas14dawes,molla15}. To compute the yields we 
integrate the mass lost from the model star during the entire stellar lifetime according to
\begin{equation}
 M_{k} = \int_{0}^{\tau}  X(k) \frac{d M}{dt} dt, \label{eq:yield}
\end{equation}
where $M_{k}$ is the yield of species $k$ (in solar masses), $dM/dt$ is the current mass-loss 
rate, $X(k)$ refers to the current mass fraction of species $k$ at the surface, and 
$\tau$ is the lifetime of the stellar model. 
The yield as expressed in Equation~\ref{eq:yield} is the amount of each element expelled into 
the interstellar medium over the stellar lifetime (in $\Msun$) and is always positive. 
In Table~\ref{tab:exampleyield} we show the first few lines of the yield tables for the 
3.5$\Msun$, $Z = 0.03$ model as an example. In Figs.~\ref{z014-yield},~\ref{z03-yield}, 
and~\ref{z007-yield} we show the stellar yields plotted against the initial stellar mass, 
for a selection of elements. In each figure we illustrate the yield and the yield weighted
by the Salpeter initial-mass function (IMF). We have chosen the Salpeter IMF for simplicity
to assess the impact of yields from low-mass AGB stars relative to those from 
intermediate-mass AGB stars of initial mass over $\approx 4\Msun$. 

In Table~\ref{tab:lithium} we present the stellar yields of Li from models with
HBB.  If column 5 is subtracted from column 6, we obtain the net yield of 
lithium, which provides an indication if the element is produced or destroyed over 
the star's lifetime.  Interestingly, all the net yields of Li from the metal-rich models 
of $Z=0.03$ are positive, even for the most massive AGB models with strong HBB. 
This is in contrast to the solar metallicity and lower metallicity $Z=0.007$ models, 
which only show positive net yields for masses near the minimum mass for HBB 
($\approx 4.5\Msun$). In more massive AGB stars of solar metallicity and lower, Li
production peaks early on, before much mass is lost from the star. By the time the superwind
begins, the star has exhausted its supply of \iso{3}He in the envelope and HBB results
in a efficient destruction of Li. 

\citet{travaglio01b} explored the Galactic chemical evolution of Li using
yields from intermediate-mass AGB stars similar to those calculated here.
The results were that intermediate-mass AGB stars do not play
a role in the chemical evolution of lithium in the Galaxy. The 
mass-loss rates of intermediate-mass stars are uncertain and a stronger mass-loss
rate that removes the envelope more quickly may well change this conclusions.
Indeed, \citet{prantzos12} concludes that a significant fraction of Li
must be produced in low and intermediate-mass. Perhaps thermohaline mixing and rotation 
play a role in the shaping the yields of Li as they do for \iso{3}He \citep{lagarde11}.

For the solar 
metallicity models, the IMF-weighted yield of N peaks at $\approx 1.5\Msun$ and at 5$\Msun$,
where the low-mass component derives from the FDU and is secondary, and the
5$\Msun$ component derives from HBB and is a mix of primary and second nitrogen. 
For Rb, the IMF-weighted yields peak at 4$\Msun$; this is because the 4$\Msun$ 
has both a \iso{13}C pocket and a burst of neutrons from the \iso{22}Ne($\alpha$,n)\iso{25}Mg reaction.
Models with the \iso{22}Ne($\alpha$,n)\iso{25}Mg reaction alone as a neutron source 
do not produce enough $s$-process elements, including Rb, to compete with production
from the \iso{13}C($\alpha$,n)\iso{16}O reaction in the lower-mass stars.
For the elements heavier than Rb, AGB stars
between about 1.5--3$\Msun$ dominate production as expected \citep[e.g.,][]{busso01}. 
The peak occurs at $\sim$ 2$\Msun$ for Ba, La, and Pb, which reflects the IMF and the TDU efficiency, 
and at 3$\Msun$ for Sr and Y, which 
reflects the contribution of the \iso{22}Ne($\alpha$,n)\iso{25}Mg neutron source to 
these elements.
Also for F the peak is 3$\Msun$ owing to the fact that the 
\iso{15}N($\alpha$,$\gamma$)\iso{19}F reaction 
that produces F in the intershell is most efficient at that mass, as discussed
earlier in \S\ref{sec:abund}.

For the metal-rich $Z=0.03$ models illustrated in Fig.~\ref{z03-yield}, 
the IMF-weighted yields of $s$-process elements
also shows a strong peak in the models that experience TDU and \iso{13}C pockets. 
The peak for all $s$-process elements is observed at the lowest
mass that becomes C-rich, which is 2.5$\Msun$ according to our assumptions, 
which reflects the shape of the IMF and also the fact that at this metallicity the 
effect of the \iso{22}Ne neutron source on elements such as Sr and Y is marginal.
The yields of light $s$-process elements (Sr, Y) are higher than the
yields of heavy $s$-process elements (Ba, La). The yields of Pb are
lower than in the models of lower metallicity. 
We conclude that AGB models with metallicity $Z=0.03$ do not contribute greatly 
to the chemical enrichment of elements heavier than La, although a full chemical
evolution model is needed to test this. The weighted yields of lighter elements C and
F show a strong increase toward lower mass, due to the IMF, while the yield of N is relatively
flat with initial mass.

The $Z=0.007$ models shown in Fig.~\ref{z007-yield} show a strong production
of all elements. Here the IMF-weighted N yields show a strong preference for intermediate-mass
AGB production owing to the efficient HBB found for models above 4.5$\Msun$. 
The effect of HBB is also visible in the yields of C and F, as a decrease in models with HBB.
The yield of Rb peaks at 4$\Msun$ as it did at solar
metallicity and for the same reasons. Elements heavier than Rb are again dominated by
the low-mass AGB stars that experience TDU and \iso{13}C pockets. As in $Z=0.014$ case, the 
Sr and Y preference for $\sim$ 3$\Msun$ models is driven by the contribution of the 
\iso{22}Ne neutron source at this mass, while the Ba, La, and Pb preference for $\sim$ 2$\Msun$ 
models is driven by the balance between the IMF and the TDU efficiency.

\begin{figure}
    \begin{center}
          \includegraphics[width=0.95\columnwidth]{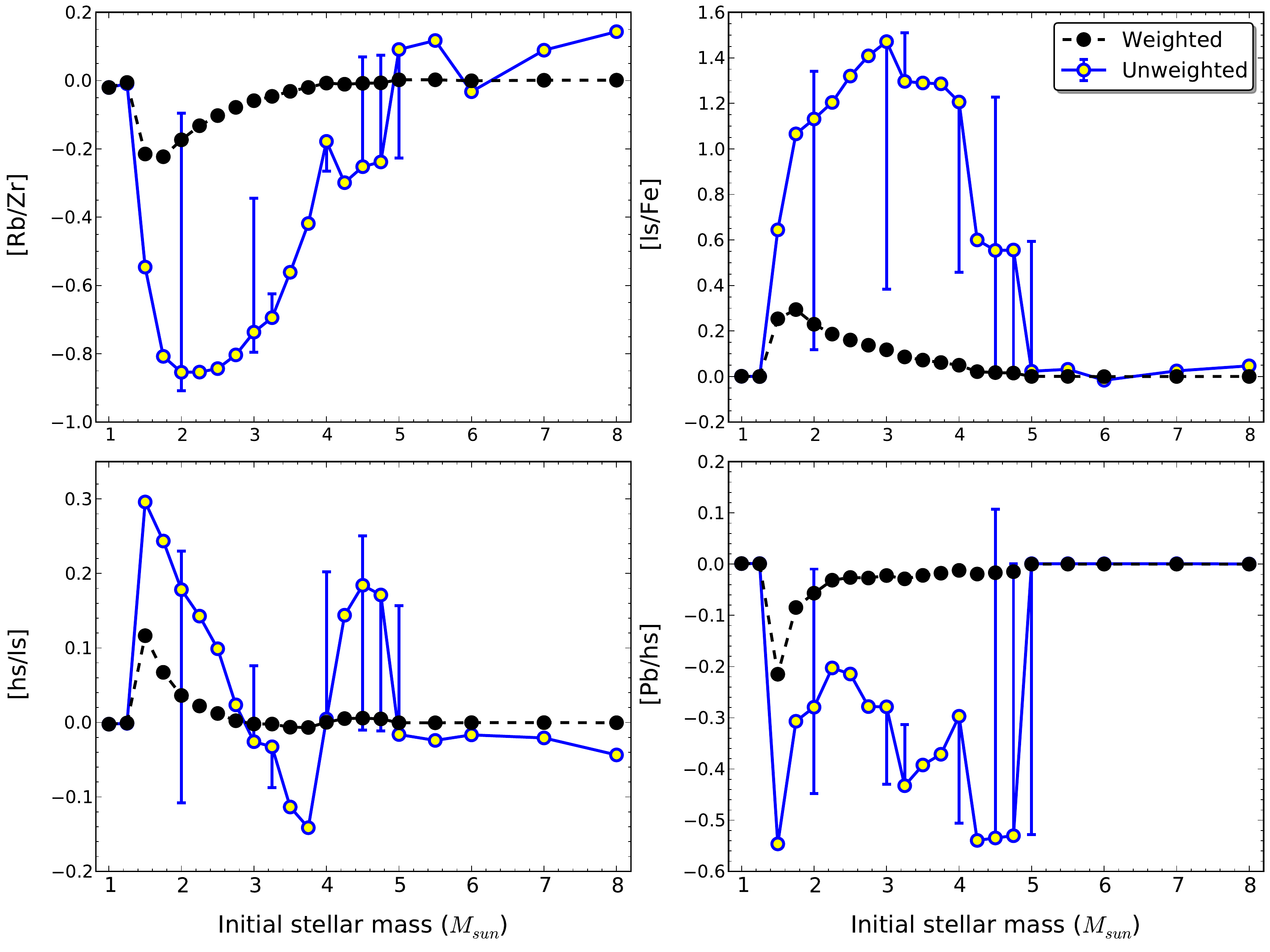}
\caption{The $s$-process indicators [Rb/Zr], [ls/Fe], [hs/ls] and [Pb/hs] as a function
of initial stellar mass (blue-yellow circles) for the $Z=0.014$ models. The ratios are 
calculated from the integrated yield abundances, not from the final surface abundances.
We varied the mass of $M_{\rm mix}$ as indicated in Table~\ref{tab:z014pmz}, with results shown 
as error bars
on the blue line.  These give some indication of the range of behaviour as a function of 
stellar mass and \iso{13}C pocket size for models of this metallicity.
The black filled circles show the $s$-process indicators weighted by the Salpeter IMF.}
    \label{z014-s-ind}
  \end{center}
\end{figure}

\begin{figure}
    \begin{center}
          \includegraphics[width=0.95\columnwidth]{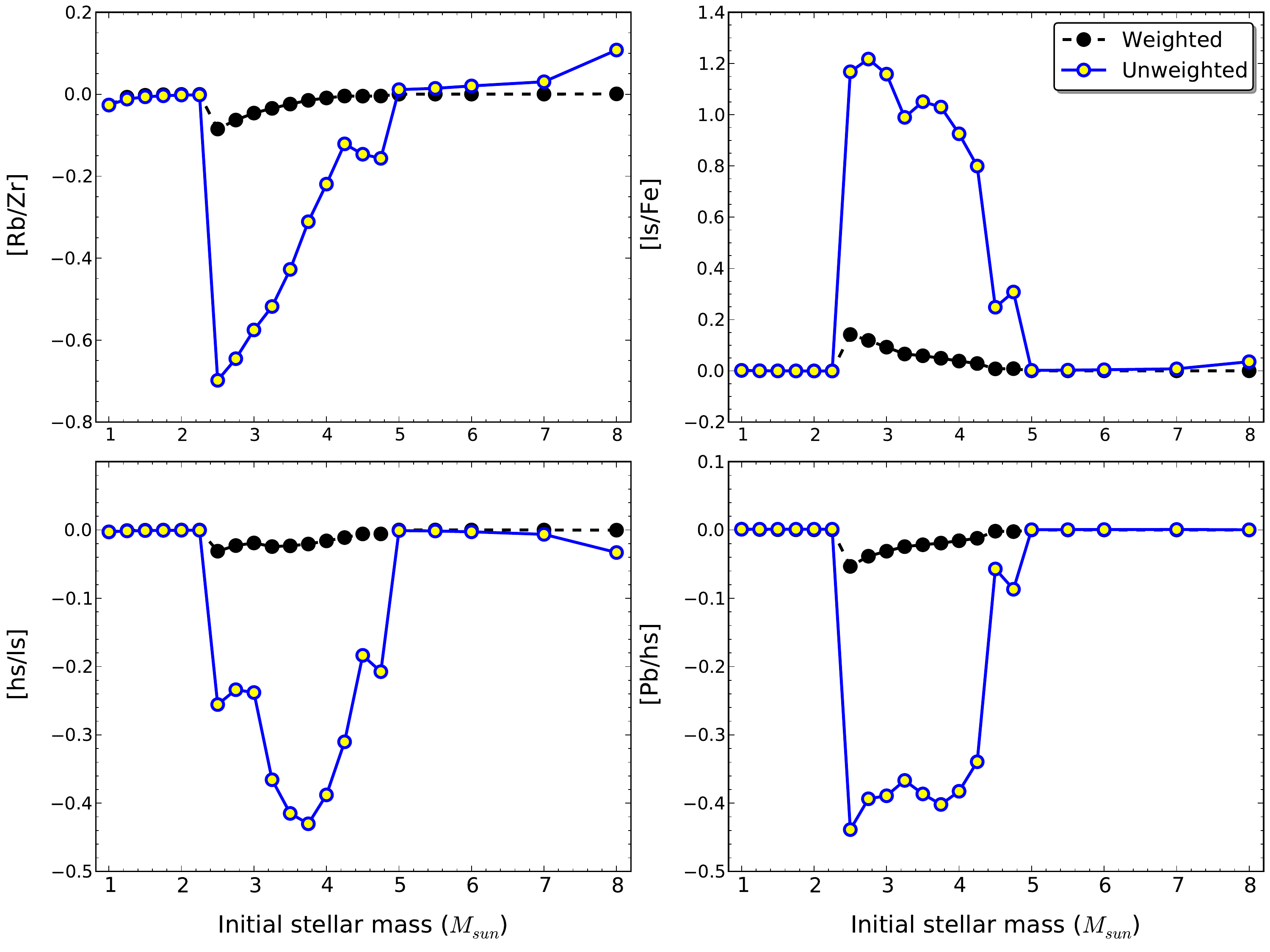}
\caption{Same as Fig.~\ref{z014-s-ind} but for the $Z=0.03$ models. Here we only show $s$-process
indicators for our standard choice of $M_{\rm mix}$.}
    \label{z03-s-ind}
  \end{center}
\end{figure}

\begin{figure}
    \begin{center}
          \includegraphics[width=0.95\columnwidth]{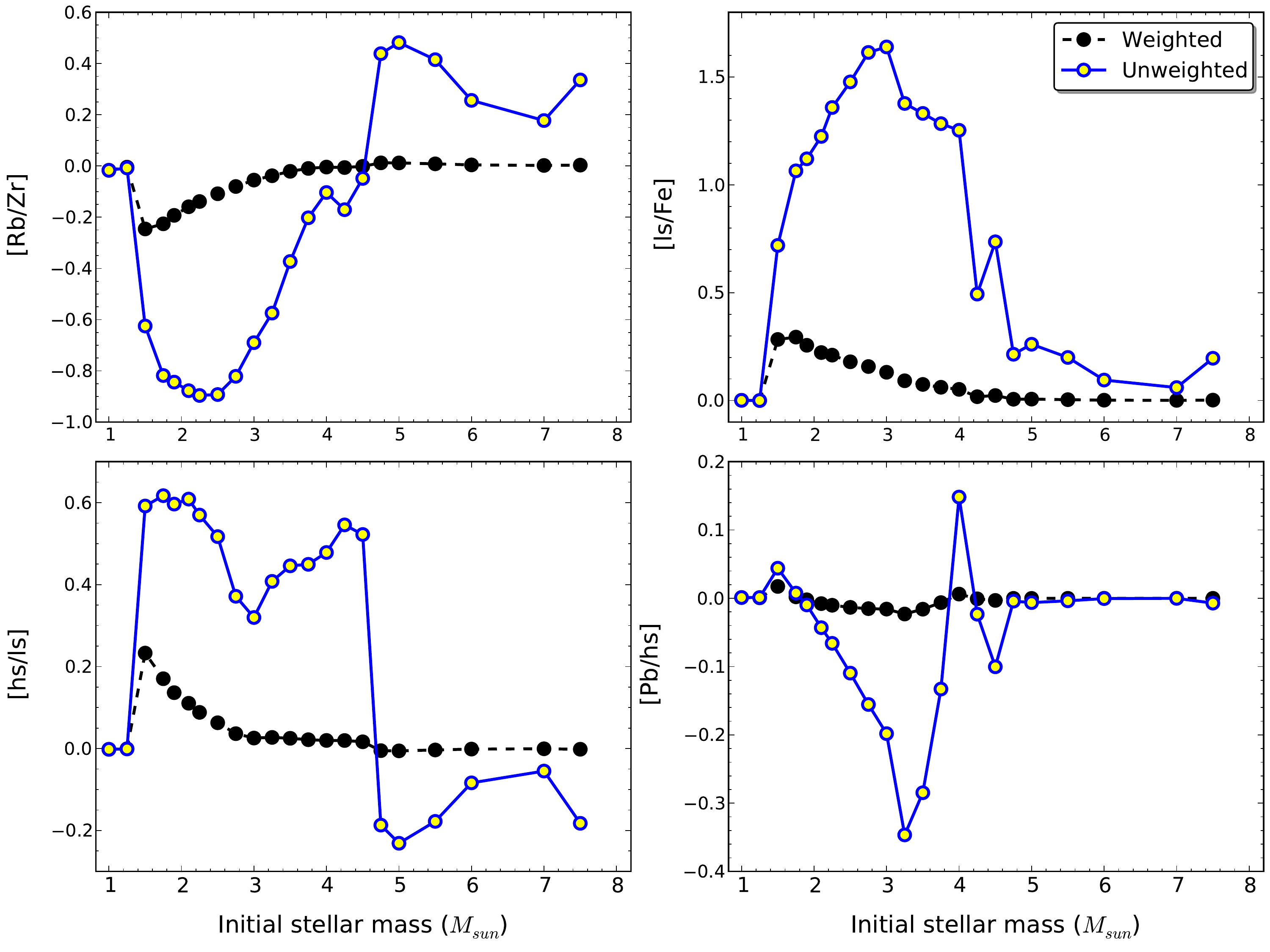}
\caption{Same as Fig.~\ref{z03-s-ind} but for the $Z=0.007$ models.}
    \label{z007-s-ind}
  \end{center}
\end{figure}

In Figs.~\ref{z014-s-ind},~\ref{z03-s-ind}, and~\ref{z007-s-ind} we show how the
$s$-process indicators [Rb/Zr], [ls/Fe], [hs/ls] and [Pb/hs] vary as a function 
of stellar mass for the three metallicities considered in this study. We also 
show the IMF-weighted $s$-process indicators. The weighting removes most of the noise and is 
indicative of what is expected in a Galactic chemical evolution model. 
In Fig~\ref{z014-s-ind} we also show the range of ratios expected from predictions
when varying the size of $M_{\rm mix}$, which controls the size of the \iso{13}C pocket
in our calculations. We only plot this for the solar-metallicity models because we
have the most extensive grid for different values of $M_{\rm mix}$ (Table~\ref{tab:z014pmz}). 
The large ranges indicate how this parameter is still the major uncertainty 
affecting the $s$-process in AGB stars, however, it cannot be considered as a real error bar, but 
rather an illustration of the difficulty of quantifying this uncertainty.

In more detail, using the [ls/Fe] ratio as a proxy of the absolute abundance produced 
(i.e., the yields) it can be seen that variations are very large when $M_{\rm mix}$ is varied 
in the case of the 2 and 3$\Msun$ stars. In these  models we varied $M_{\rm mix}$ down to zero, 
in which case there are almost no neutrons. The error bars cover the observations of AGB 
stars that range from 0 to 1 dex \citep{busso01,abia02}, however, we need to be cautious  
before reaching conclusions on the size and variation of $M_{\rm mix}$, because low [ls/Fe] 
values can also be explained by stars of lower mass, as shown in the figure. 

The models 
mostly affected by the uncertainty on $M_{\rm mix}$ are those at the transition between the 
low and the intermediate-mass regime, i.e., between 4 and 5$\Msun$ in the figure. 
It is difficult to identify clear observational constraints attributable to this mass 
range. Moreover, model uncertainties will play a role in determining exactly how $M_{\rm mix}$ 
should scale down as the intershell mass becomes smaller with evolution.
Also HBB can be partially activated  during the AGB in models of this mass, which
means that the effect of the hot dredge-up \citep{goriely04} may appear or disappear.
For these models we experiment with a range of plausible $M_{\rm mix}$ values. For higher 
masses, the uncertainty becomes smaller given the observational evidence \citep{garcia13} 
for the absence of \iso{13}C pockets.

The [Rb/Zr] ratio is very sensitive to the neutron source active in the He-intershell: 
negative values indicate that the \iso{13}C($\alpha$,n)\iso{16}O neutron 
source reaction is dominant while positive values indicate that the \iso{22}Ne 
neutron source. This comes about because the density of neutrons released by the 
\iso{13}C reaction are lower ($n_{\rm n} \lesssim 10^{8}$ n/cm$^{3}$) than the peak 
neutron density from the \iso{22}Ne neutron source 
\citep[$n_{\rm n} \approx 10^{13}$ n/cm$^{3}$ which allows branching points open to 
produce Rb, e.g., ][]{vanraai12,fishlock14b}.
We only see positive values for the intermediate-mass models above 4$\Msun$ (unless 
a \iso{13}C pocket is included), with the largest [Rb/Zr] obtained in the lowest 
metallicity models. 

\section{Comparison with other yield sets} \label{sec:compare}

Here we compare the results from the 1.5, 3 and 6$\Msun$ models to 
models of the same mass and similar $Z$ from the FRUITY database \citep{cristallo11,cristallo15}. 
We also compare our 5$\Msun$, $Z=0.014$ model and yields to models of the same mass
from the NuGrid database \citep{pignatari13}. We note that a good agreement
between yields does not mean that the yields are free of uncertainties, indeed, such
agreement could be coincidental. Comparing to observations is the most reliable method
to verify the accuracy of the predictions.

\citet{cristallo15} also use $Z=0.014$ for the solar metallicity, which makes a direct
comparison straightforward.  There are no AGB models published with $Z=0.03$, 
although there are slightly super-solar metallicity ($Z=0.02$) models available 
from the FRUITY database. FRUITY include models with $Z=0.008$ or $Z=0.006$, and
our lower metallicity $Z=0.007$ models are in the middle of that range. 

For the comparison it is useful to keep in mind the differences between the
FRUITY models and those presented here. In particular, we assume no mass loss
on the RGB and \citet{vw93} mass-loss on the AGB. \citet{cristallo15} 
adopt a Reimer's mass-loss rate with $\eta = 0.4$ for the pre-AGB phase
and then use the formulae from \citet{straniero06} for the AGB.
In additional to differences in the mass-loss rates, there are also significant 
differences in our treatment of convection and convective borders, which were 
discussed in Section~\ref{sec:c13pocket}. Differences in the nuclear physics 
input can also sometimes play a role.

\begin{figure}
    \begin{center}
          \includegraphics[width=0.95\columnwidth]{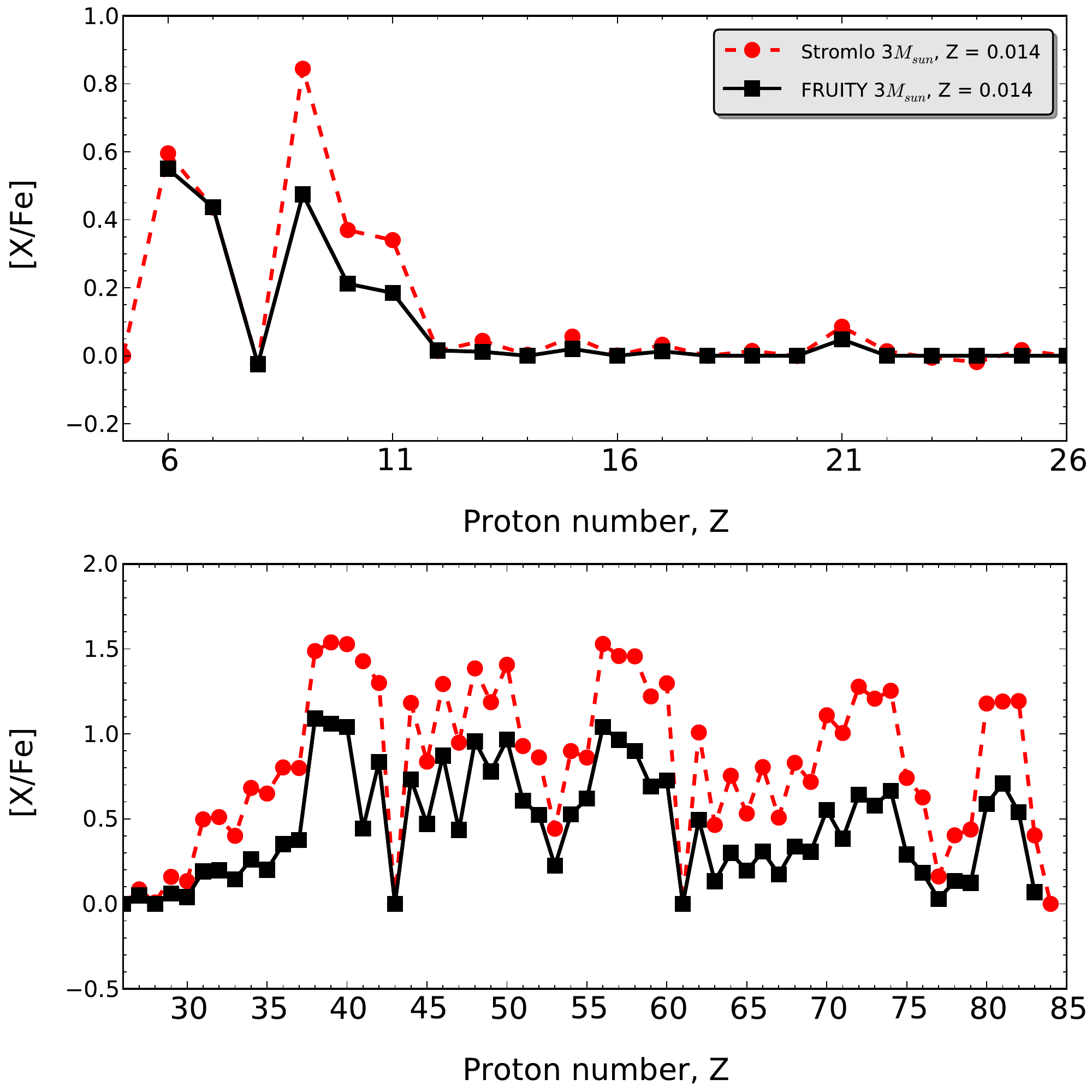}
\caption{Comparison between the nucleosynthesis results from 3$\Msun$, $Z= 0.014$ models.
We are showing the final elemental surface composition (in [X/Fe]) for each case.
The Stromlo model refers to the model presented here. We have set [Tc, Pm/Fe] = 0 in the figures
(Tc and Pm have proton numbers $z = 43$ and $z = 61$, respectively), whereas in reality these 
abundance ratios are not defined, because these elements are radioactive 
and have a zero solar abundance.}
    \label{m3z014-compare}
  \end{center}
\end{figure}

In Fig.~\ref{m3z014-compare} we compare the results for the 3$\Msun$, $Z =0.014$ models. The 
agreement between the light elements C, N and O is reasonably good, while our model produces 
more F, Ne and Na, and more elements heavier than Fe. The $s$-process distribution is very similar 
with the [Ba/Sr] and [Ba/Pb] ratios the same to within 0.15 dex. Here we are comparing to our
3$\Msun$, $Z = 0.014$ model with $M_{\rm mix} = 2 \times 10^{-3}\Msun$, noting that the
models with smaller partially mixed zones produce less heavy elements, and are closer to the 
FRUITY model. However,
even the model with the smallest  $M_{\rm mix} = 1 \times 10^{-4}\Msun$ still produces 
[F/Fe] = 0.74, which is higher than the FRUITY predictions for this mass ([F/Fe] = 0.475),
although the agreement with Ne and Na improves. This suggests that our F abundances
are higher owing to differences in the nuclear network. 

Comparing the amount of TDU, the Stromlo 3$\Msun$ and 6$\Msun$, $Z=0.014$ models 
dredge-up roughly the same amount of material ($\approx 0.1\Msun$), as shown in 
Fig.~\ref{mdredge}. The 6$\Msun$, $Z=0.007$ also dredges up about $0.1\Msun$. The 
FRUITY 3$\Msun$ of solar composition dredges up 0.06$\Msun$, roughly 60\% less material 
than the Stromlo model of the same mass and composition. This difference explains 
the higher absolute F, Na, and $s$-process abundances that we see in 
Fig.~\ref{m3z014-compare}.  The Stromlo model dredges up more 
material because it experiences more TPs and deeper TDU.
The FRUITY model has 17 thermal pulses compared to 28 in the Stromlo case.
While the FRUITY model experiences deeper TDU sooner than the Stromlo case, the
Stromlo model has a higher peak TDU efficiency as measured using the parameter $\lambda$ 
(c.f., $\lambda_{\rm max} \approx 0.8$ compared to $\lambda_{\rm max} \approx 0.6$ 
from the FRUITY model). Interestingly, the C abundances are similar in the 3$\Msun$ 
models, even though the Stromlo model dredges up 60\% more He-shell. That the surface C 
abundances are coincidently the same implies that the FRUITY models have a higher C 
abundance in their intershell. The cause probably lies in the choice of triple-$\alpha$ 
reaction rate: FRUITY use the NACRE rate, which is 10\% faster (at T=200 MK) than the 
rate we use from the JINA reaclib database.

\begin{figure}
    \begin{center}
          \includegraphics[width=0.95\columnwidth]{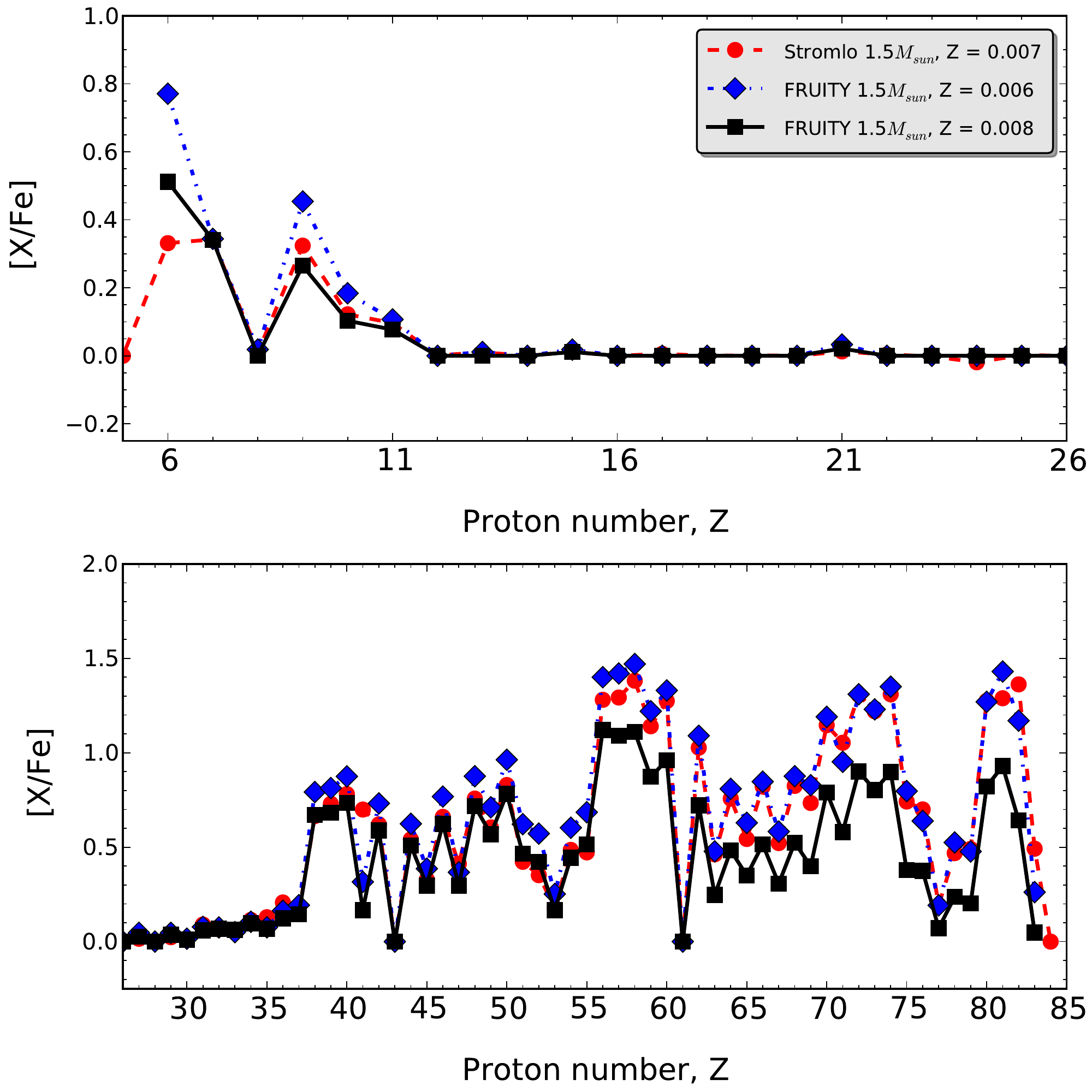}
\caption{Comparison between the nucleosynthesis results from 1.5$\Msun$, $Z= 0.007$ Stromlo
model and the 1.5$\Msun$ models from FRUITY with $Z=0.006$ and $Z=0.008$. Results are shown
for the final elemental surface composition.}
    \label{m1.5z007-compare}
  \end{center}
\end{figure}

\begin{figure}
    \begin{center}
          \includegraphics[width=0.95\columnwidth]{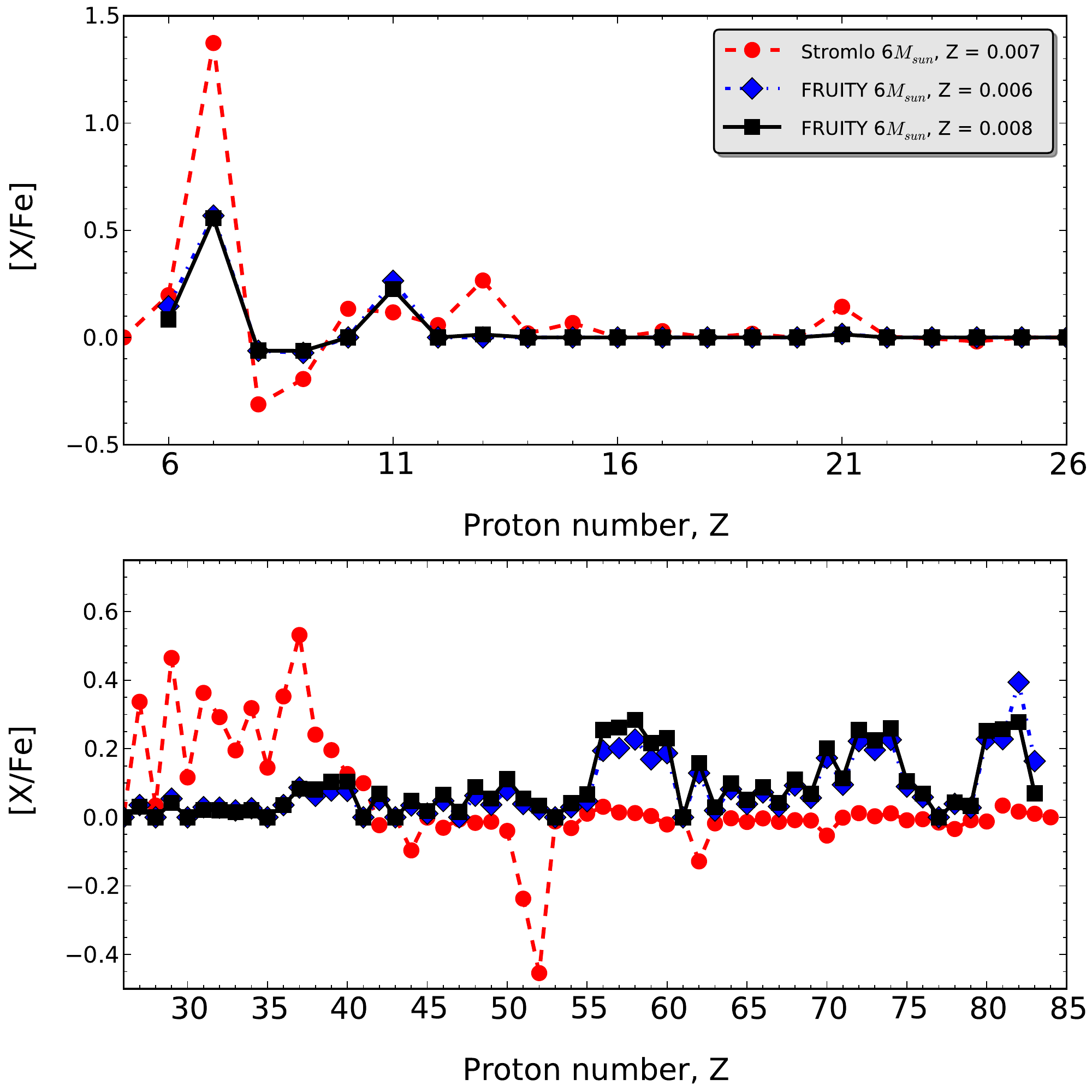}
\caption{Same as Fig.~\ref{m1.5z007-compare} except for the 6$\Msun$, $Z = 0.007$ model.}
    \label{m6z007-compare}
  \end{center}
\end{figure}

In Figs.~\ref{m1.5z007-compare} and~\ref{m6z007-compare} we compare the results from
our 1.5$\Msun$ and 6$\Msun$ models with $Z= 0.007$ to models of similar metallicity from
the FRUITY database. For the 1.5$\Msun$ case, the two FRUITY models of $Z=0.006$ and 
$Z=0.008$ predict C and F abundances that increase with decreasing $Z$. The Stromlo 
1.5$\Msun$ model produces less C than either FRUITY model but similar F, confirming 
the results for the 3$\Msun$, solar metallicity case. The results for the heavy elements 
are similar for all three models, with the Stromlo model lying close to the predictions 
of the FRUITY $Z=0.006$ model. The main reason for the reasonable agreement between
the distribution of abundances shown in Fig.~\ref{m1.5z007-compare} is that the models 
dredge-up roughly about the same amount of material. 
The Stromlo 1.5$\Msun$, $Z = 0.007$ model dredges up $7.6\times 10^{-3}\Msun$ 
compared to $\approx 8\times 10^{-3}\Msun$ and $0.014\Msun$ from the FRUITY 
$Z=0.008$ and $Z=0.006$ models, respectively.

The largest discrepancies are found for models of intermediate mass such as the 
case of the 6$\Msun$, $Z=0.007$ model illustrated in Fig.~\ref{m6z007-compare}.
The Stromlo model has stronger HBB, which is evident from the production of N and Al
and the destruction of O and F. On the other hand, there is very little destruction 
of these elements in the 
FRUITY 6$\Msun$ case. The discrepancies between the FRUITY models and
the Stromlo model continue into the heavy elements. The Stromlo model shows evidence
of the \iso{22}Ne($\alpha$,n)\iso{25}Mg neutron source, with a strong production of elements at
the first $s$-process peak (the final [Rb/Fe] = 0.53) and little $s$-process 
production beyond that\footnote{The low abundance of Te is because we
do not include all stable isotopes of this element; see discussion in \citet{lugaro12}.}.
In contrast, the FRUITY 6$\Msun$ models show almost no production of light $s$-process 
elements and instead some production around the Ba and Pb peaks, caused by
the formation of a small \iso{13}C pocket \citep{straniero14,cristallo15}.
We do not see any indication of activation of the \iso{22}Ne in the FRUITY model
from the surface composition. 

The discrepancies illustrated in Fig.~\ref{m6z007-compare} can be mostly 
traced back to the input physics used in the models. The FRUITY 6$\Msun$ models 
loses considerably more mass before the first TP and experience fewer TPs,
as discussed above for the 1.5$\Msun$ and 3$\Msun$ models. Our 6$\Msun$, $Z=0.007$
model shown in Fig.~\ref{m6z007-compare} has 64 TPs, roughly  three times more TPs
in the FRUITY models of similar composition. 
Fewer TPs results in less material dredged to the stellar surface, where 
the FRUITY models of 6$\Msun$, $Z=0.006$ and $Z=0.008$ dredge up a factor of
11 to 18 times less than our $6\Msun$, $Z = 0.007$ model.  Less material 
from the He-intershell means that there is fewer \iso{12}C nuclei
in the envelope available to be converted into \iso{14}N by HBB. 

The smaller dredge-up in the FRUITY models explains why our model
has higher absolute enhancements in $s$-process elements. However,
this alone does not explain the differences in the $s$-process distributions. Our
model shows activation of the \iso{22}Ne($\alpha$,n)\iso{25}Mg 
neutron source, with a peak in production around Rb. We do not include 
\iso{13}C pockets into the 6$\Msun$ models, so we see no evidence of the 
\iso{13}C neutron source, that is, we have low Ba, La and Pb abundances. 
In contrast, the FRUITY 6$\Msun$ models only show enhancements
in Ba and heavier elements, indicating that their models develop \iso{13}C
pockets as a consequence of their treatment of convection and
convective borders \citep[as discussed in][]{cristallo15}. On the other hand, 
the signature of the \iso{22}Ne($\alpha$,n)\iso{25}Mg neutron source 
is not present at the stellar surface. This again can probably be traced
back to the different mass-loss law. The \iso{22}Ne($\alpha$,n)\iso{25}Mg 
neutron source is typically activated more efficiently in the later 
rather than earlier TPs (opposite to the \iso{13}C neutron source, which is 
activated from the first TDU episode), which explains why the FRUITY models 
do not show its effect at the stellar surface.

Observations of Galactic AGB stars \citet{garcia13} found no evidence 
for the activation of the \iso{13}C neutron source in intermediate-mass 
stars, when using Tc as a tracer. On the other hand, observations of 
the brightest intermediate-mass AGB stars both in the Galaxy and in the 
Magellanic Clouds reveal a surface chemistry that is O-rich and $s$-process rich
\citep{wood83,garcia06,garcia09}. This is at odds with the model predictions from 
the FRUITY database which predict a final [Rb/Fe] $<0.1$, although the Stromlo 
models also fail to quantitatively account for the huge Rb enrichments 
observed \citep[e.g.,][]{vanraai12,karakas12}. Circumstellar effects have been 
found to  cause large overestimates when determining the abundances of Rb.
These effects do not remove the enrichments and in fact bring the model predictions 
by \citet{karakas12} more in line with observations \citep{zamora14}.

The other major discrepancy between the intermediate-mass models concerns the
strength of HBB.  Other stellar evolution codes predict HBB as strong as ours 
\citep[e.g.][]{herwig04a,weiss09,pignatari13} or stronger, in the case of the 
Full Spectrum of Turbulence models calculated using the ATON code 
\citep[e.g.,][]{ventura13}.  
In comparison to the FRUITY models, the Stromlo code predicts considerably 
higher temperatures at the base of the envelope during HBB 
\citep[as also highlighted by][]{fishlock14b,shingles15}.
From Fig.~9 from \citet{cristallo15} the peak HBB temperatures in their 
6$\Msun$, $Z=0.014$, $Z = 0.008$ and $Z=0.006$ models are 
$\log T_{\rm bce}^{\rm max}/K \approx 7.25$, 7.35, and 
7.45, respectively\footnote{these numbers are approximate and read from
the figure. HBB temperatures are not provided on the on-line FRUITY 
database.}. In contrast, our 6$\Msun$ models have maximums of 
$\log T_{\rm bce}^{\rm max}/K = 7.85$, 7.93, 7.96 for metallicities $Z=0.03$,
$Z= 0.014$, and $Z=0.007$, respectively.  \citet{cristallo15} explore
various reasons why their models experience lower temperatures at the
base of the convective envelope but did not identify the cause. 

We initially speculated if the higher mass loss experienced by the
FRUITY models is the cause. The 6$\Msun$, $Z=0.014$ model from 
\citet{cristallo15} loses $\approx 1\Msun$
on the early AGB and enters the TP-AGB with a total mass of 5.1$\Msun$, 
effectively a 5$\Msun$ model star. This is very different to what we find
when using the \citet{vw93} prescription, where less than 0.1$\Msun$ is lost
during the early AGB. The reason for the rapid early AGB mass loss comes
down to the calculation of the pulsation period, which determines the
rate of mass loss. A detailed comparison
between our 6$\Msun$ model and the FRUITY model
finds excellent agreement between the radius and luminosity at the beginning
of the early AGB and at the start of the TP-AGB\footnote{we also find 
excellent agreement between our core H and He-burning lifetimes to better
than 5\%.}.
However our pulsation periods, as calculated using the period-radius-mass
relation from \citet[][their Eq.~4]{vw93}, are roughly a factor of 10
lower. \citet{cristallo15} calculate the pulsation period using a
$M_{\rm k}-\log P$ relation \citep[e.g.,][]{whitelock08}, which requires
calculation of $M_{\rm k}$ first from stellar variables.  

Going back to the connection between HBB and mass loss, we perform a test
calculation and adopt the faster \citet{bloecker95} mass-loss rate 
on the AGB with $\eta = 0.4$ in a 6$\Msun$, $Z=0.014$ model in order to
strip of the envelope mass quickly. Now the mass-loss increases
such that we only calculate 10 TPs (c.f., 53 when using \citet{vw93}). 
The peak temperature still reaches $\log T_{\rm bce}^{\rm max}/K = 7.72$, 
which is roughly a factor of 3 higher than the peak HBB temperature found 
in the Cristallo et al. model of the same mass and composition. We conclude 
that while the mass-loss rates found by the FRUITY calculations are higher, 
the lower HBB temperatures are not caused by their choice of mass loss 
on the AGB. 

The process of HBB in intermediate-mass AGB models explains the observational 
fact that the most luminous AGB population in the Magellanic Clouds is dominated 
by O-rich AGB stars, while the less luminous AGB population is dominated by 
C-rich AGB stars \citep{wood83}. \citet{cristallo15} note that the effect of 
HBB may be mimicked by rapid rotation in intermediate-mass AGB stars. 
At the present time there are no intermediate-mass AGB models with
rotation available at the metallicities of the Large Magellanic Cloud or 
solar metallicity to check if rapid rotation does indeed mimic the signature 
of HBB.

The brightest O-rich AGB stars in the Magellanic Clouds and Galaxy are also rich in Li 
\citep{smith90b,garcia13}. In Table~\ref{tab:lithium} we show the Li abundances
from our models with HBB, which confirms that our intermediate-mass AGB models with 
HBB become Li-rich \citep[e.g., see also][]{lattanzio97,vanraai12}.
For example, the peak Li abundances in models of 6$\Msun$ are $\log \epsilon$(Li) = 4.87, 
4.28, 4.42, respectively, for $Z=0.03, 0.014$, and $Z=0.007$, confirming 
that the models are both O-rich and super-Li rich, at least for a while. 
It would be interesting to test if intermediate-mass AGB models with rotation 
also become Li rich. 

The NuGrid/{\tt MESA} collaboration calculated models and yields of low and intermediate-mass
AGB stars \citep{pignatari13}. Their grid includes models of $M=1.65, 2, 3, 4, 5\Msun$
with $Z=0.01$ and $Z=0.02$. For each ($M, Z$) combination stellar evolution model data 
and detailed yields are provided. Given the discrepancies we found between our 
intermediate-mass AGB models with HBB and those from the FRUITY database we  
compare our results to the NuGrid intermediate-mass 5$\Msun$ AGB models. Our 5$\Msun$, 
$Z=0.014$ model in the middle of NuGrid metallicity range. First,
we note that the NuGrid 5$\Msun$ models both experience HBB, with peak temperatures
at the base of the envelope of 50~MK and 65~MK, respectively 
\citep[Tables 6 and 7 from][]{pignatari13}. At first glance our 5$\Msun$, $Z =0.014$ model
would appear to experience stronger HBB, with a peak temperature of 75~MK. However, this 
is likely because the \citet{vw93} mass-loss rate is slower and maintains 
a more massive envelope for longer, compared to the \citet{bloecker95} mass-loss rate
used in the {\tt MESA} calculations. 
This can be quantified by comparing the number of thermal pulses: our model has 41~TPs
compared to the 25 TPs and 22 TPs experienced by the 5$\Msun$ NuGrid models of 
$Z=0.02$ and $Z=0.01$, respectively. Note that the NuGrid 5$\Msun$ models reach their peak
HBB temperature after about 11~TPs. After 11~TPs the temperature at the base of the 
envelope of our 5$\Msun$ model is 52~MK, comparable to the NuGrid $Z=0.02$ model. 
This demonstrates that HBB is consistent between our models and the 
NuGrid/{\tt MESA} models.

The Stromlo model dredges up roughly 3 and 4 times as much material as the NuGrid $Z=0.02$ and 
$Z=0.01$ models, respectively. This, combined with envelope burning extended over more TPs, 
results in higher yields of most elements, except for carbon, which is destroyed in the 
Stromlo models. 
The yield of Rb is the same to within 5\% between the 5$\Msun$, $Z=0.01$ and the Stromlo
model with a small \iso{13}C pocket ($M_{\rm mix} = 1\times 10^{-4}\Msun$), while 
yields of other heavier $s$-process elements are higher in the Stromlo case. Examples include
Zr (factor 4 higher), and Ba/La (roughly factor of 40 higher), and Pb (factor 2.8 higher).
That the yields of Rb are similar is a coincidence: the higher TDU offsets the slower rate 
for the \iso{22}Ne($\alpha$,n)\iso{25}Mg reaction that we adopt. The
{\tt MESA} calculations adopt the faster NACRE rate for this important neutron producing
reaction; \citet{karakas12} showed that this rate increases Rb production.
Here we adopt the slower \citet{iliadis10} rate. The Stromlo 5$\Msun$ model without 
a \iso{13}C pocket produces fewer heavy elements than the NuGrid model by almost a factor of
two in most cases. This indicates two things: the importance of small \iso{13}C pockets 
in intermediate-mass AGB models and that the NuGrid 5$\Msun$ models have small \iso{13}C 
pockets as a result of the convective boundary mixing scheme employed in the {\tt MESA} 
evolutionary calculations  \citep[see discussion in][]{pignatari13}.  In summary,
the NuGrid/{\tt MESA} models are qualitatively similar to ours: HBB occurs and produces N, heavy
element production occurs and is dominated by the light $s$-process elements around Rb,
even in the presence of a small \iso{13}C pocket.

\section{Conclusions} \label{sec:conclude}

In this study we presented surface abundances for elements and isotopes, as well as elemental
stellar yields for an extensive set of AGB models covering a large range in mass for three
metallicities, and including the first
nucleosynthesis predictions and yields for twice-solar metallicity AGB models. 
For solar metallicity and models with a super-solar metallicity of $Z=0.03$, 
we include models from 1$\Msun$ to the C-O core limit of 8$\Msun$, noting that the limiting
mass at these metallicities will end as a hybrid CO(Ne) white dwarf. For the lower metallicity
models of $Z=0.007$, we include models from 1$\Msun$ to 7.5$\Msun$, where 7$\Msun$ is the C-O
core limit and the 7.5$\Msun$ is a super-AGB model. This is one of the few predictions of 
heavy element production for a super-AGB star in the literature. 

We include a fine grid of masses which allows us to observe behaviour in the surface abundances
and yields that would otherwise be missed. An example is the prediction that models between 
about 4.25$\Msun$ and 4.75$\Msun$ with solar metallicity that experience
HBB have low \iso{12}C/\iso{13}C ratios $<10$ but become C-rich. 
A fine grid of masses is required in order to provide accurate yield predictions for
Galactic chemical evolution models, as discussed by \citet{izzard04b} in the context of 
synthetic, rapid AGB models. 

We find that the surface abundances and yields of the super-solar metallicity AGB stars
of $Z=0.03$ are different to their lower metallicity counterparts. Only models above 2.5$\Msun$
experience TDU and only models above 5$\Msun$ experience HBB. We find a narrow range of
carbon star production, between 2.5$\Msun$ and 4$\Msun$, even when we include convective overshoot.
Without overshoot, the range decreases to 3.25--4$\Msun$, as discussed in \citet{karakas14b}.
This mass range also dictates which masses produce $s$-process elements, as shown in
\S\ref{sec:abund} and \S\ref{sec:yields}. Super-solar metallicity models of intermediate-mass
are predicted to produce only light elements from HBB and no heavy elements.
Interestingly these models are all net Li producers, including
the most massive models of 8$\Msun$ near the CO-core mass limit which experience 
efficient HBB. This is in contrast to models of lower metallicity and solar composition,
which only have positive net lithium production near the minimum stellar mass for 
HBB $\approx 4.5\Msun$. Higher mass models destroy Li by the end of the AGB phase.

We weight the yields by a Salpeter IMF to show how the yields and $s$-process abundances
change in a stellar population; this gives an indication how important each mass range is
for chemical evolution.  As found elsewhere  \citep{travaglio04, bisterzo14} the yields 
from the intermediate-mass AGB stars do not play an important role at these metallicities, 
except for N and Rb.  Production of a substantial yield of Rb from AGB stars requires the 
contribution from stars of $\approx 4\Msun$ that include both
the \iso{13}C and \iso{22}Ne neutron sources. Models with the \iso{22}Ne
source alone are not predicted to play a major role in the production of Rb or other heavy
elements in the Galaxy, at least at these metallicities.

Finally, we compare our models to the FRUITY nucleosynthesis predictions from \citet{cristallo15}
and the NuGrid AGB models and yields from \citet{pignatari13}.
We find reasonable qualitative agreement between the FRUITY models of low-mass AGB stars of 
1.5$\Msun$ and 3$\Msun$ for all metallicities (except $Z=0.03$, for which no FRUITY models 
are available to compare). However, we find large discrepancies between the FRUITY 
6$\Msun$ intermediate-mass models and ours, while our 5$\Msun$ model
is qualitatively similar to the 5$\Msun$ NuGrid models.  While this is a problem 
that may be resolved through a careful comparison with observations, the yields of these 
intermediate-mass AGB stars are not important for bulk Galactic chemical evolution studies. There 
may be places in the Galaxy, however, where the yields of intermediate-mass AGB stars are important 
and these possibly include Galactic globular clusters \citep{ventura13,shingles14,straniero14}. One 
issue with models of intermediate-mass AGB stars is the paucity of observations, which are 
required to confront theoretical predictions. Future observations of bright intermediate-mass AGB 
stars in stellar populations of varying metallicity would be an invaluable resource in furthering 
our understanding of these enigmatic objects.

\acknowledgments

The authors thank the Referee for encouraging us to dig deeper into the differences between
our models and other published results.  
We would also like to thank Marco Pignatari for providing information about the NuGrid AGB models and yields,
and Sergio Cristallo for providing detailed information about the FRUITY models.
This research was undertaken with the assistance of resources from the National Computational 
Infrastructure (NCI), which is supported by the Australian Government.   AIK was supported 
through an Australian Research Council Future Fellowship (FT110100475). ML is a Momentum 
(``Lend\"ulet-2014'' Programme) project leader of the Hungarian Academy of Sciences.

\appendix

\section{Appendix material}

Examples of each of the data table types are included in the Appendix. 

\begin{table}
\renewcommand{\arraystretch}{1.05}
 \begin{center}
  \caption{Example of the surface abundance tables available. We show the first few lines at the
beginning of the 3.5$\Msun$, $Z = 0.03$ model table, and the first few lines after thermal pulse \#15.}
 \label{tab:examplesurf}
  \vspace{1mm}
   \begin{tabular}{ccccccc}
   \tableline\tableline
\multicolumn{7}{l}{\# }\\
\multicolumn{7}{l}{\# Initial mass = 3.500, Z =  0.0300, Y =  0.300, $M_{\rm mix}$ = 1.00E-03}\\
\multicolumn{7}{l}{\# }\\
\multicolumn{7}{l}{\# Initial abundances} \\
\#El & $Z$  & $\log e(X)$ & [X/H] & [X/Fe] & [X/O] & $X(i)$ \\
\multicolumn{7}{l}{...}\\
\multicolumn{7}{l}{\# TP \hspace{0.2mm}  Mass   \hspace{3mm}   Mcore   \hspace{3mm}  Menv   \hspace{3mm}  log L} \\ 
\multicolumn{7}{l}{\# 15 3.498070 0.694067 2.804000 4.096260} \\
 \#El & $Z$  & $\log e(X)$ & [X/H] & [X/Fe] & [X/O] & $X(i)$ \\ 
  p & 1 & 12.000000 & 0.000000 & 0.000000 & 0.000000 & 6.59291E-01 \\
 he & 2 & 11.050125 & 0.152777 & -0.187120 & -0.167322 & 3.16795E-01 \\
  c & 6 & 8.799475 & 0.218580 & -0.121316 & -0.101519 & 3.83509E-03 \\
  n & 7 & 8.199430 & 0.774327 & 0.434431 & 0.454228 & 4.03913E-03 \\
  o & 8 & 9.059475 & 0.320489 & -0.019408 & 0.000000 & 1.17543E-02 \\
  f & 9 & 4.789476 & 0.370921 & 0.031024 & 0.050432 & 8.41870E-07 \\
...  \\
\multicolumn{7}{l}{\# Elemental abundance ratios:} \\
\multicolumn{7}{l}{\# He/H =  1.2100E-01, C/O =  4.3460E-01, N/O = 3.92498E-01} \\
\tableline
\tableline
  \end{tabular} 
\\
 \end{center}
\end{table}

\begin{table}
\renewcommand{\arraystretch}{1.00}
 \begin{center}
  \caption{Example of the isotopic abundance tables available. We show the first few lines of the 
3.5$\Msun$, $Z = 0.03$ model table for the first five isotopic ratios in the table.}
 \label{tab:exampleiso}
  \vspace{1mm}
   \begin{tabular}{cccccc}
   \tableline\tableline
\multicolumn{6}{l}{\# }\\
\multicolumn{6}{l}{\# Initial mass = 3.500, Z =  0.0300, Y =  0.300, $M_{\rm mix}$ = 1.00E-03}\\
\multicolumn{6}{l}{\# }\\
\multicolumn{6}{l}{\#Initial isotopic abundance ratios:}\\
\# c12/c13 &  n14/n15 &  o16/o17 &  o16/o18 & mg24/mg25 & ... \\
 8.940E+01 & 4.476E+02 & 2.632E+03 & 4.988E+02 & 7.899E+00 & ... \\
\multicolumn{6}{l}{\# }\\
\multicolumn{6}{l}{\# During TP-AGB}\\
\multicolumn{6}{l}{\# }\\
  1.963E+01 & 2.757E+03 & 3.911E+02 & 6.987E+02 & 7.957E+00 ... \\
  1.963E+01 & 2.757E+03 & 3.911E+02 & 6.987E+02 & 7.957E+00 ... \\ 
  1.964E+01 & 2.754E+03 & 3.913E+02 & 6.986E+02 & 7.957E+00 ... \\
...  \\
\tableline
\tableline
  \end{tabular} 
\\
 \end{center}
\end{table}

\begin{table}
\renewcommand{\arraystretch}{1.05}
 \begin{center}
  \caption{Example of the yield tables available. We show the first few lines of the 
3.5$\Msun$, $Z = 0.03$ model yield table.}
 \label{tab:exampleyield}
  \vspace{1mm}
   \begin{tabular}{ccccccc}
   \tableline\tableline
\multicolumn{7}{l}{\# Initial mass = 3.500, Z =  0.0300, Y =  0.300, $M_{\rm mix}$ = 1.00E-03} \\
\multicolumn{7}{l}{\# Final mass = 0.727, Mass expelled =   2.7730} \\
 \#El & $Z$  & $\log e(X)$ & [X/H] & [X/Fe] & $X(i)$ & Mass($i$) \\
  p &  1 & 12.000000 & 0.000000 & 0.000000 & 6.35230E-01 & 1.76149E+00 \\
 he &  2 & 11.119857 & 0.189857 & $-$0.165502 & 3.32440E-01 & 9.21855E-01 \\
  c &  6 &  9.147092 & 0.677092 & 0.321733 & 1.06204E-02 & 2.94504E-02 \\
  n &  7 &  8.643233 & 0.773233 & 0.417875 & 3.88194E-03 & 1.07646E-02 \\
  o &  8 &  9.057708 & 0.327708 & $-$0.027650 & 1.15152E-02 & 3.19316E-02 \\
  f &  9 &  5.442862 & 0.982862 & 0.627503 & 3.31925E-06 & 9.20429E-06 \\
\multicolumn{7}{l}{...} \\
\tableline \tableline
  \end{tabular} 
\\
 \end{center}
\end{table}

\bibliographystyle{apj}
\bibliography{apj-jour,library}


\end{document}